\documentclass[12pt,preprint]{aastex}

\usepackage{natbib}

\shorttitle{Infrared Power-law Galaxies in the CDFS}
\shortauthors{A. Alonso-Herrero et al.}

\begin{document}


\title{Infrared power-law galaxies in the Chandra Deep 
Field South: AGN and ULIRGs}



\author{A. Alonso-Herrero\altaffilmark{1,2}, 
P. G. P\'erez-Gonz\'alez\altaffilmark{2}, 
D. M. Alexander\altaffilmark{3},
G. H. Rieke\altaffilmark{2},
D. Rigopoulou\altaffilmark{4},
E. Le Floc'h\altaffilmark{2}, 
P. Barmby\altaffilmark{5}, 
C. Papovich\altaffilmark{2}, 
J. R. Rigby\altaffilmark{2}, 
F. E. Bauer\altaffilmark{6},
W. N. Brandt\altaffilmark{7}, 
E. Egami\altaffilmark{2}, 
S. P. Willner\altaffilmark{5},
H. Dole\altaffilmark{8,2},  
J.-S. Huang\altaffilmark{5}
}

\altaffiltext{1}{Departamento de Astrof\'{\i}sica Molecular 
e Infrarroja, Instituto de Estructura de la Materia, CSIC, E-28006 Madrid, Spain; 
e-mail: aalonso@damir.iem.csic.es}
\altaffiltext{2}{Steward Observatory, The University of Arizona, 
933 N. Cherry, Tucson, AZ 85721}
\altaffiltext{3}{Institute of Astronomy, University of Cambridge, 
Madingley Road, Cambridge, CB3 0HA, UK}
\altaffiltext{4}{Department of Astrophysics, Oxford University, 
Keble Rd, Oxford, OX1 3RH, UK}
\altaffiltext{5}{Harvard-Smithsonian Center for Astrophysics, 
Cambridge, MA 02138}
\altaffiltext{6}{Columbia Astrophysics Laboratory, Columbia University
             Pupin Laboratories, 550 W. 120th St., Rm 1418, NY, 10027}
\altaffiltext{7}{Department of Astronomy and Astrophysics; The Pennsylvania
State University; 525 Davey Lab; University Park, PA 16802}
\altaffiltext{8}{Institut d'Astrophysique Spatiale, b\^at 121, Universit\'e
Paris Sud, F-91405 Orsay Cedex, France}



\begin{abstract}
We investigate the nature of a sample of 92 {\it Spitzer}/MIPS 
$24\,\mu$m selected galaxies in the CDFS,    
showing power law-like emission in the Spitzer/IRAC 
3.6--8$\mu$m bands. The main goal is to determine whether the galaxies 
 not detected in X-rays 
($47\%$ of the sample) are part of the hypothetical population  of 
obscured AGN not detected even in deep X-ray surveys.
The majority of the IR power-law
galaxies are ULIRGs at $z>1$, and those 
with LIRG-like IR luminosities are usually detected in X-rays. 
The optical to IR spectral energy distributions
(SEDs) of the X-ray detected galaxies are almost equally
divided between a BLAGN SED class (similar to an optically selected QSO) 
and a NLAGN SED (similar to the BLAGN SED 
but with an obscured UV/optical continuum). 
A small fraction of SEDs resemble warm 
ULIRG galaxies (e.g., Mrk~231). Most galaxies
not detected in X-rays have SEDs in the NLAGN+ULIRG class as they tend
to be optically fainter, and possibly more obscured.  
Moreover, the IR power-law galaxies have SEDs significantly 
different from those of high-$z$ ($z_{\rm sp} >1$) IR
($24\,\mu$m) selected and optically bright 
(VVDS $I_{\rm AB} \le 24$) 
star-forming galaxies whose SEDs show a very prominent stellar
bump at $1.6\,\mu$m. 
The galaxies detected in X-rays have $2-8\,$keV rest-frame luminosities
typical of AGN. The galaxies not detected in X-rays have global 
X-ray to mid-IR SED properties that make them good candidates to contain IR
bright X-ray absorbed AGN. If all these sources are actually obscured
AGN, we would observe a ratio of obscured to unobscured 24~$\mu$m
detected AGN of 2:1, whereas models predict a ratio of up to
3:1. Additional studies using {\it Spitzer} to detect X-ray-quiet AGN
are likely to find more such obscured sources.
\end{abstract}




%
\section{Introduction}
Active galactic nuclei (AGN) are sources of luminous X-ray emission, and 
at cosmological distances AGN are routinely selected from deep X-ray 
($<10\,$keV, {\it Chandra} and 
{\it Newton-XMM)} exposures (see Brandt \& Hasinger 2005 for a recent
review). 
However, obscured (Compton thick) AGN are
thought to be a major contributor
to the hard X-ray background (Comastri et al. 1995) and the majority of 
them might not be detected in these X-ray surveys. 
Since a large fraction of their soft X-ray, 
UV and optical emission is absorbed, and presumably reradiated in the infrared
(IR), obscured AGN are also believed to 
make a significant contribution to the IR and submillimeter
backgrounds (Fabian \& Iwasawa 1999; Almaini, Lawrence, \& Boyle 1999). 
Fadda et al. (2002) used ISOCAM on the {\it Infrared Space Observatory}
({\it ISO)} in X-ray survey fields to estimate that X-ray emitting AGN contribute 
up to 17\% of the $15\,\micron$ extragalactic background (see also
Alexander et al. 2002; Silva, Maiolino, \& Granato 2004). 

Since their discovery, it has been recognized that the UV and 
optical continuum of luminous QSOs could be described with a power 
law that can continue  all the way into the near, mid, and even far-IR 
(see e.g., Neugebauer et al. 1979; Elvis et al. 
1994). This behavior often arises from the combination of a number of
source components that broaden the purely stellar spectral energy distribution (SED),
rather than from a dominant contribution from a true power-law source (see, e.g.,
Rieke \& Lebofsky 1981). 
Nonetheless, it is a surprisingly general characteristic. 
At lower luminosities, for 
instance X-ray selected Seyfert 1 galaxies, 
the most common SED
from the UV to the far-IR is a relatively flat $\nu f_\nu$ distribution, although
for some of the sources the observed SEDs show the effects of dust 
obscuration and thermal emission (e.g., Ward et al. 1987; Carleton et al.
1987). A similar result is found for optically selected Seyfert galaxies 
in the CfA sample, although Seyfert 2 galaxies tend to have 
steeper IR slopes, probably due to the effects of dust emission/obscuration (e.g., 
Edelson, Malkan, \& Rieke 1987). Optical and IR power-law emission
is  also present in X-ray selected QSOs (HEAO, Kuraszkiewicz  et al. 2003) 
and near-IR selected  QSOs (2MASS, Wilkes et al. 2005).

Therefore, sensitive measurements in the IR range provide an opportunity to look for 
obscured AGN not identified in X-ray and optical surveys.
First, unlike normal and star forming
galaxies whose near-IR SEDs are dominated by the $1.6\,\mu$m stellar bump, AGN dominated 
galaxies (even those obscured at optical wavelengths) could be potentially 
identified through power-law emission in the rest-frame 
near-IR ($1-5\,\mu$m). For instance, Egami et al. (2004) have found 
SCUBA/VLA sources not detected in X-rays whose SEDs show a power-law shape,
and probably contain an AGN.
Second, Alexander et al. (2005)  have shown that 
among the most IR-luminous high-$z$ galaxies --- those detected 
at high redshift with SCUBA ---
the AGN fraction is high, at least 40\%, suggesting that 
the most IR-luminous high-$z$ galaxies may be good candidates to host AGN. 
Very recent {\it Spitzer} spectroscopy of IR luminous galaxies at $z\simeq
1-2$ has shown a similar fraction of galaxies containing an AGN
(Yan et al. 2005).

In this paper we investigate the nature of galaxies selected at the 
{\it Spitzer}/MIPS $24\,\mu$m band and within the {\it Chandra} 
Deep Field South (CDFS), and that show 
power-law-like emission in the {\it Spitzer}/IRAC $3.6-8\,\mu$m bands.  By
selecting galaxies at $24\,\mu$m we ensure that at $z>1$ a large fraction
of them will be in the ultraluminous IR galaxy (ULIRG, $L_{\rm IR} >
10^{12}\,{\rm L}_\odot$\footnote{The IR luminosities are 
in the $8-1000\,\mu$m range (Sanders \& Mirabel 1996).}) 
class (see P\'erez-Gonz\'alez et al. 2005).
Throughout this paper we use $H_0 = 71\,{\rm km \,s}^{-1}\,{\rm Mpc}^{-1}$, 
$\Omega_{\rm M} = 0.3$, and $\Omega_\Lambda = 0.7$.

\section{Spitzer  Observations}

We have obtained  $24\,\mu$m and $70\,\mu$m MIPS (Rieke et al. 
2004) observations 
of the CDFS covering a total area of $1.5\deg \times 0.5\deg$. In this work we 
will only focus on the field of view (FOV) with {\it Chandra} coverage
($\simeq 390$ sqr arcmin, 
Giacconi et al. 2002; Alexander et al. 2003a). The MIPS data were reduced using the 
Data Analysis Tool (DAT) package developed by 
the MIPS instrument team (Gordon et al. 2005).

The $24\,\mu$m source detection and photometry, described in  detail by 
Papovich et al. (2004) and P\'erez-Gonz\'alez et al. (2005), assumed that all
sources were point-like, given the $\simeq 5.8\arcsec$ FWHM angular resolution.
PSF-fitting with DAOPHOT (Stetson 1987) packages  
{\it daofind, phot}, and {\it allstar} in {\sc iraf} gave positions, fluxes, and
uncertainty estimates.
The final 5$\sigma$ flux density limit was $83\,\mu$Jy and the catalog is 80\%
complete at this level (see Papovich et al. 2004).
We measured photometry at $70\,\mu$m using the $24\,\mu$m positions as a priori  
information.  We used a version of the DAOPHOT/ALLFRAME package (Stetson  
1994), which simultaneously fits the $70\,\mu$m PSF at the  
position of each $24\,\mu$m source in the $70\,\mu$m image for sources within  
50\arcsec \ of nearby object centroids.  This method has the advantage of  
recovering photometry from faint sources in the $70\,\mu$m image, even  
when sources are partially confused.    The errors on the  
$70\,\mu$m sources include an uncertainty on the fitting of sources with  
overlapping isophotes, providing an estimate of the error on each  
source owing any source confusion.
The DAOPHOT/ALLSTAR PSF-fitting software
detected $70\,\mu$m sources down to an upper limit of $3\,$mJy, fainter
than that of Dole et al. (2004) because of the use of the $24\,\mu$m a
priori information.

The CDFS was observed by IRAC (Fazio et al. 2004) 
at 3.6, 4.5, 5.8, and $8\,\micron$ covering an area of 
$1.0\deg \times 0.5\deg$. IRAC photometry began with the 
Basic Calibrated Data images from the
{\it Spitzer} Science Center pipeline, version 9.5.  Custom IDL software
projected the individual frames onto a uniform grid of 0\farcs6
pixels, conserving flux and rejecting outliers such as cosmic ray
hits. Source
detection and photometry were carried out with {\sc sextractor} (Bertin \&
Arnouts 1996). We used
a small circular aperture ($3\arcsec$ in diameter) and an aperture
correction to get the total flux (assumed to be the one corresponding to
a circular aperture of diameter 24.4$\arcsec$) of each source. The
aperture corrections were calculated from in-flight PSFs. This 
procedure is similar to that used by Huang et al. (2004). The 
photometry errors take into account the IRAC photometric calibration
uncertainty and the {\sc sextractor} photometry errors.

\section{Selection of IR power-law galaxies}
Galaxies dominated by AGN emission typically show a power law
($f_\nu \propto \nu^\alpha$, where $\alpha$ is the 
spectral index) SED at rest-frame optical, near- and mid-IR
wavelengths, though with a variety of slopes.
Elvis et al. (1994) showed that optically selected 
QSOs  have average optical to IR spectral indices around 
$\alpha=-1$ (see also Neugebauer et al. 1979). More recent results 
from the Sloan Digital 
Sky Survey (SDSS) show that the optical spectral indices of 
optically selected QSOs are
in the range of $\alpha= -0.5$ to
$-2$ (Ivezi\'c et al. 2002, their figure 17). Of course this
does not necessarily mean that the spectral index is the same in the
near- and mid-IR. 2MASS QSOs also show power-law-like 
SEDs in the near- and mid-IR, very similar
to those of optically selected QSOs, although the UV-optical continuum shape is
steeper than that of optically and X-ray selected QSOs (Wilkes et al. 2005).
ULIRGs ($L_{\rm IR} > 10^{12}\,{\rm L}_\odot$) containing an AGN, in 
particular warm ULIRGs, also have near and mid-IR SEDs 
resembling power laws (see Sanders et al.
1988, and more recently Klaas et al.  2001).  Some of these warm ULIRGs in the 
local Universe are also
identified with optically selected QSOs, and all of them are known
to host an AGN. Moreover, warm ULIRGs
have been proposed as an intermediate stage in the evolution between
ULIRGs and QSOs (Sanders et al. 1988; Sanders et al. 2004, 
and references therein).  Thus the presence of 
an IR power law in LIRGs and ULIRGs could also be an indication 
of the presence of an AGN.

To construct an initial power-law galaxy sample, we started with 
the $24\,\mu$m source catalog (Papovich et al. 2004) 
in the FOV of the X-ray {\it Chandra} observations of the CDFS. 
We cross-correlated the positions of these $24\,\mu$m sources 
with sources detected in the four IRAC bands using a 1.5\arcsec \ search
radius, and 
constructed observed $3.5-8\,\mu$m SEDs, requiring that the sources be
detected in all four IRAC bands. We fitted 
the SEDs with a power law $f_\nu \propto \nu^\alpha$ between $3.6$ and 
$8\,\mu$m, and minimized $\chi^2$ to select 
galaxies whose IRAC SEDs followed a power law with 
spectral index $\alpha < -0.5$. These objects formed the candidate list
for our sample, subject to construction of full SEDs toward shorter
wavelengths. Further tests of the power-law nature of these objects were used
to derive the final sample, as described in the Appendix.

Recently  Lacy et al. (2004)
and Stern et al. (2005)  have used IRAC observations of SDSS QSOs in the 
{\it Spitzer} First Look Survey and spectroscopically confirmed AGN in the 
Bo\"otes Field respectively,
 to derive an empirical color selection for AGN. They show that bright
QSOs and broad-line AGN 
occupy a very distinct region in the IRAC color-color diagram. 
Follow-up optical spectroscopy of some of the 
optically faint AGN candidates identified by Lacy et al. (2004) 
show high ionization narrow lines typical of obscured type 2 AGN (see Lacy 
et al. 2005) demonstrating the effectiveness of this method. We note that our criteria 
select galaxies along a line in the IRAC color-color diagram, 
rather than using the full region identified by Lacy et al. 2004 
(their figure~1). Our approach is therefore more restrictive than theirs,
since we depend entirely on the SED behavior to identify candidate AGN.
Barmby et al. (2005) present a detailed comparison between
the IRAC power-law and the IRAC color-color selection methods for AGN in the Extended
Groth Strip.


The final sample consists of 92
galaxies referred to here as ``IR power-law galaxies''. They have been
selected without regard to X-ray properties, and should be unbiased
in that respect, and can therefore be used to probe the number of X-ray quiet
AGN in the CDFS. Thirty-seven members of the sample 
are also detected with MIPS at $70\,\mu$m down
to a limit of $f_{70\mu{\rm m}} \simeq 3\,$mJy. 
Table~1 gives  {\it Spitzer} coordinates, flux densities and
uncertainties, as well as the fitted IRAC spectral indices $\alpha$ for the
sample of IR power-law galaxies.

An interesting aspect of these galaxies is that we found a maximally steep
optical-to-IR slope corresponding to $\alpha \sim -2.8$ (see Table~1). 
A similar limit was found by
Rigby et al. (2005a) in studying a complete sample of optically faint X-ray sources in the 
CDFS. Stickel et al. (1996) studied a complete sample of faint identifications
of flat spectrum radio sources. They also found that the steepest
optical-IR SEDs were characterized by $\alpha \sim -2.5$. This
slope therefore seems to define a limiting behavior for AGN SEDs.

\section{X-ray Observations}
The CDFS field has ultra-deep 1\,Ms {\it Chandra} observations (Giacconi
et~al. 2002 and Alexander et~al. 2003a), providing a sensitive census
of AGN activity. The aim-point sensitivities in the
$0.5-8.0\,$keV (full), $0.5-2.0\,$keV (soft), and 
$2-8\,$keV (hard) bands are
$\approx 1.3\times10^{-16}$~erg~cm$^{-2}$~s$^{-1}$, $\approx
5.2\times10^{-17}$~erg~cm$^{-2}$~s$^{-1}$, and $\approx
2.8\times10^{-16}$~erg~cm$^{-2}$~s$^{-1}$, respectively. As an
example, assuming an X-ray spectral slope of $\Gamma=$~2.0, a source
detected with a flux of $\approx
1.0\times10^{-16}$~erg~cm$^{-2}$~s$^{-1}$ would have both observed and
rest-frame luminosities of $\approx 5.8\times 10^{41}$~erg~s$^{-1}$,
$\approx 8.4\times 10^{42}$~erg~s$^{-1}$, and $\approx 2.8\times
10^{43}$~erg~s$^{-1}$ at $z=1$, $z=3$, and $z=5$, respectively.

The majority of the X-ray data are taken from 
tables~3a and 3b of the main catalog of Alexander et al. (2003a). However, 
for the faintest X-ray
sources, the data were taken from a largely unpublished supplementary
catalog. We cross-correlated the positions of our 92 IR power-law SED galaxies
with the X-ray sources using a $1.2-3$\arcsec \ search radius, with the latter for
X-ray sources at large off-axis angles.  
Apart from the sources in common with the Alexander et al. (2003a)
catalog, there is a further source in the Giacconi et al. (2002) catalog. For this source we 
converted the $2-10\,$keV band to the $2-8\,$keV and $0.5-8\,$keV bands assuming $\Gamma=1.4$.
We found that 49 galaxies are detected at least 
in one of the {\it Chandra} bands, whereas 43 galaxies ($47\%$) 
are not detected  in any of the {\it Chandra} 
bands. For the X-ray undetected sources we calculated 3$\sigma$ upper limits
(assuming $\Gamma = 1.4$) 
in the full, soft, and hard bands following Alexander et al. (2003a). 
The X-ray information on the IR power-law galaxy sample is given in Table~2,
including the off-axis angle that shows the sensitivity variation 
across the X-ray field. At off-axis angles of less than $9\arcmin$ (where the
sensitivity is more uniform), the
fraction of sources not detected at any {\it Chandra} band is $41\%$.

Fig.~1 compares the X-ray and mid-IR emission of the CDFS galaxies detected both at
$24\,\mu$m and in hard X-rays  (filled dots) from Alonso-Herrero et al.
(2004) and Rigby et al. (2004), and among them those identified as IR
power-law  
SED galaxies (squares). The IR power-law SED galaxies comprise about one third of 
all the hard X-ray sources with $24\,\mu$m detections. We also show those 
IR power-law SED galaxies 
with only upper limits in the {\it Chandra} hard X-ray band ($2-8\,$keV). 
The IR and 
X-ray properties of our sample will discussed in more detail in \S6 and
\S7, respectively.

\begin{figure}
\epsscale{0.8}
\hspace{4cm}
\plotone{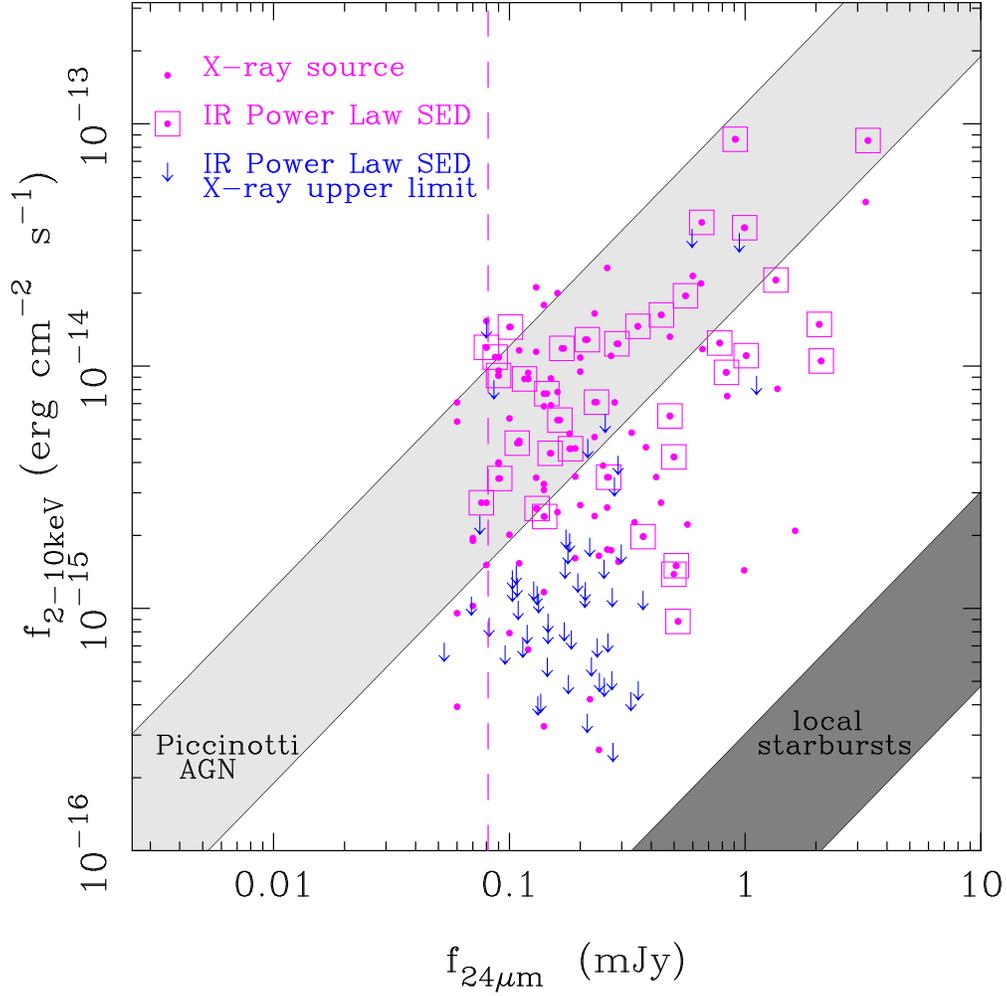}
\caption{$24\,\mu$m fluxes vs. $2-10\,$keV 
X-ray fluxes of X-ray selected sources with $24\,\mu$m
counterparts in the CDFS (dots, see Rigby et al. 2004); those
also marked with a square symbol are IR power-law SED galaxies in our sample. 
Those IR power-law SED galaxies in our sample not detected in 
hard X-rays are shown as upper limits (arrows).
The {\it Chandra} $2-8\,$keV fluxes have been converted 
to $2-10\,$keV fluxes assuming a power law with photon
index $\Gamma = 1.4$, for easy comparison with figure~1 in Alonso-Herrero et
al. (2004).
The dashed line is the 80\% completeness limit for 
the  CDFS  $24\,\mu$m source catalog (Papovich et al. 2004). 
The lightly shaded area is the extrapolation of the median hard X-ray to mid-IR 
ratios ($\pm 1\sigma$) of local ($z<0.12$) hard X-ray selected
AGN (Piccinotti et al. 1982) with mid-IR emission. 
The dark shaded area is the extrapolation of local starburst galaxies from 
Ranalli, Comastri, \& Setti (2003).}
\vspace{0.4cm}
\end{figure}

\begin{figure*}
\epsscale{0.9}
\plottwo{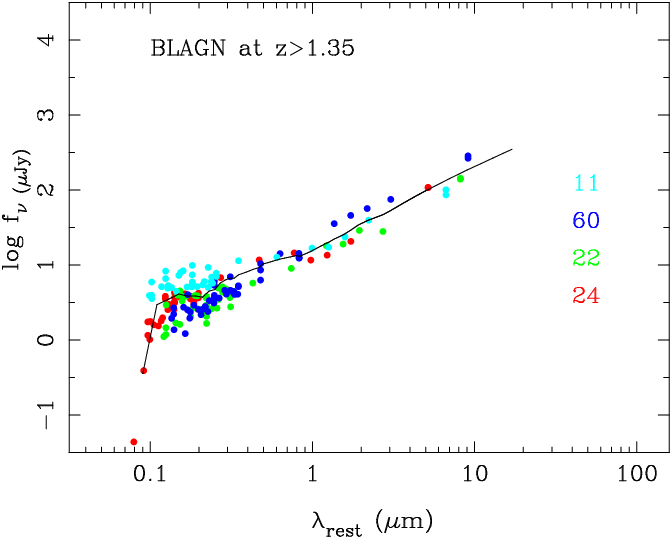}{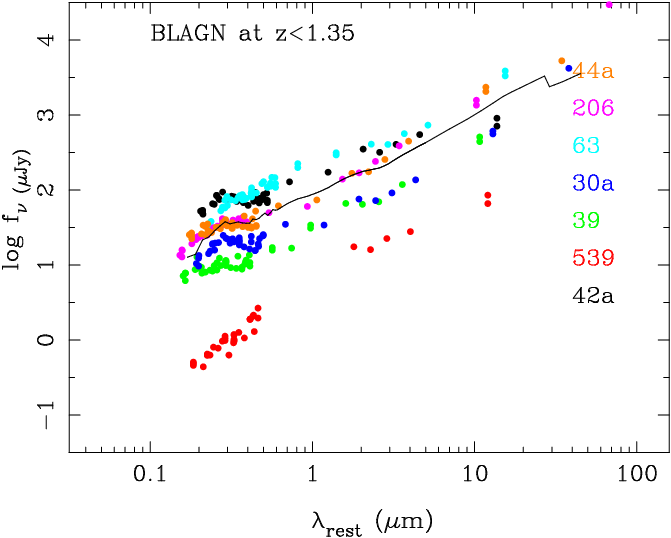}

\plottwo{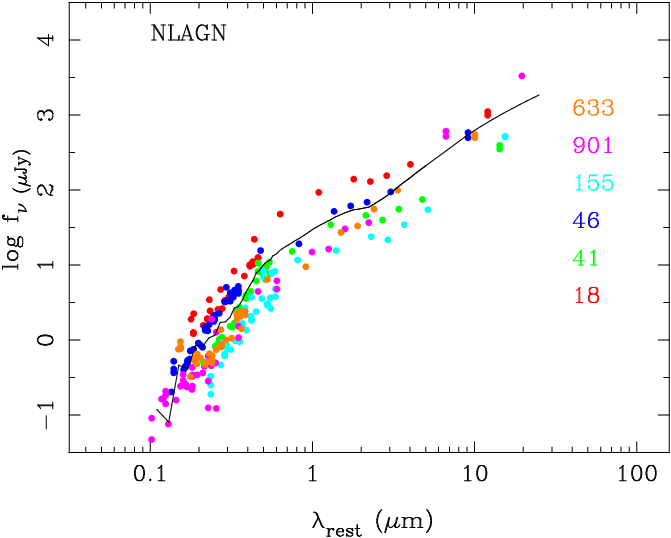}{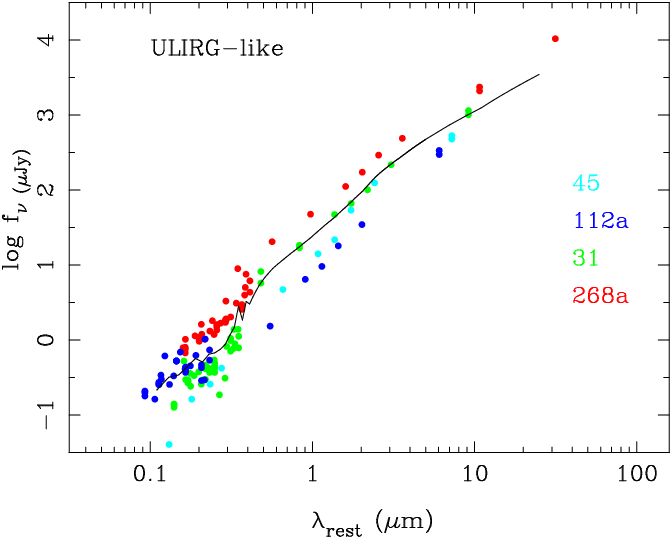}
\caption{Rest-frame SEDs (dots) of IR power-law galaxies detected 
in X-rays for 21 galaxies with spectroscopic redshifts 
available from the Szokoly et
al. (2004) sample. The 
galaxy names shown on the right hand side of each panel 
refer to source IDs in Szokoly et al. (2004). For each 
type of galaxy we have constructed an average template, shown as
the solid line in each panel.}
\vspace{0.4cm}
\end{figure*}

\section{Spectral Energy Distributions of IR power-law galaxies}

\subsection{X-ray detected galaxies with spectroscopic redshifts}
We cross-correlated our sample against the spectroscopic redshift catalogs
of Szokoly et al. (2004) and the VIRMOS-VLT Deep Survey (VVDS, 
Le F\`evre et al. 2004) for CDFS galaxies. 
Twenty-three IR power-law galaxies detected in X-rays have spectroscopic 
redshifts, of which twenty-two are in the
Szokoly et al. (2004) sample and thus an activity class based on  emission
lines is also available. The Szokoly et al. (2004) activity classes are: 
broad-line AGN (BLAGN), high 
excitation line galaxies (HEX), and low excitation line
galaxies  (LEX). 

We can use the Szokoly et al. (2004) measurements to determine the efficiency of our
criterion in finding  X-ray QSOs. 
Of the twenty-nine CDFS sources classified as X-ray QSOs 
($L_{\rm 0.5-10keV}> 10^{44}\,{\rm erg \, s}^{-1}$) by Szokoly et al. 
(2004), twenty-three are detected at $24\,\mu$m.
Our IR power-law  criteria identifies eleven out of the fourteen broad-line 
QSOs, plus two narrow-line unabsorbed QSOs. 
Out of the nine X-ray type-2 QSOs 
in Szokoly et al. (2004)  detected at $24\,\mu$m, 
we only select three as IR power-law galaxies. Of the six 
X-ray type-2 QSOs not identified with our sample selection criteria, 
three are missing an IRAC band, and the other three do not show a power-law
SED (i.e., they appear dominated by stellar emission). 

There are six further IR power-law galaxies in common with the Szokoly et al. (2004) 
sample. They are classified as AGN in terms of their X-ray luminosities ($L_{\rm 0.5-10keV} =
10^{42}- 10^{44}\,{\rm erg \, s}^{-1}$), with two broad-line AGN, and the
other four with narrow lines (HEX and LEX classes). 
We note that there are no galaxies with 
absorption lines (the ABS class of Szokoly et al. 2004) included in our
sample of IR power-law galaxies, 
as those galaxies tend to be dominated by stellar emission 
(Rigby et al. 2005b). 

In Fig.~2 we show the rest-frame SEDs of the galaxies in our sample detected
in X-rays and with spectroscopic redshifts. We define 
three SED categories:

\begin{itemize}
\item
The broad-line AGN (BLAGN) SED category includes  galaxies whose SEDs resemble those of 
optically selected QSOs, such as the median QSO of Elvis et al. (1994), 
and of X-ray selected QSOs (Kuraszkiewicz et al. 2003)
--- that is, galaxies with an optical to mid-IR continuum 
almost flat in $\nu f_\nu$ with a UV bump. All but one of the galaxies 
in the BLAGN SED category 
are X-ray QSOs in terms of their X-ray luminosities  (Szokoly et al. 2004),
and
all of them show broad lines. We have further 
divided the BLAGN galaxies in our sample into two arbitrary redshift intervals
to determine if there are significant differences in their SEDs. 

\item
 The second SED class consists of galaxies with IR SEDs similar to the 
BLAGN SED galaxies but whose UV and optical continua are much steeper
(obscured) and resemble the SEDs of 2MASS QSOs
(Wilkes et al. 2005). The galaxies will be referred to as 
narrow line AGN (NLAGN) since in terms of the 
optical spectroscopic classification most of these
galaxies do not show broad lines. Using the classifications of 
Szokoly et al. (2004), we find three HEX galaxies, two LEX galaxies, 
and one BLAGN. 

\item
The last SED class is for galaxies
with the steepest IR SEDs, resembling that of the local warm ULIRG 
Mrk~231 (see Fig.~2). Although there is no formal division between this SED class and
the NLAGN SED class, the ULIRG SED galaxies do not show 
a prominent $1.6\,\mu$m stellar bump. In terms of their X-ray 
luminosities, three are 
classified by Szokoly et al. (2004) as QSOs, 
and the other one is a type 2 AGN. In terms of
their spectroscopic classification (Szokoly et al. 2004), 
there are three HEX galaxies and one 
LEX galaxy.  
\end{itemize}

For each of the SED classes defined above, we have constructed
typical SEDs that can be used as templates to classify the
galaxies. These templates used a number of techniques such
as averaging the observations of a number of galaxies to increase
the signal and to smooth over individual eccentricities. As
Fig.~2 shows, the templates match the observations of individual
galaxies well. We note that the stellar bump at $1.6\,\mu$m is present, 
although it is not very prominent, on the average templates of NLAGN
and low-$z$ BLAGN. The average templates of the high and low-$z$ BLAGN SED are very
similar except that in the high-$z$ BLAGN AGN the $1.6\,\mu$m stellar bump
seems to be absent (note however, that only four galaxies were used to
construct this average template).

\begin{figure*}
\epsscale{0.9}
\plotone{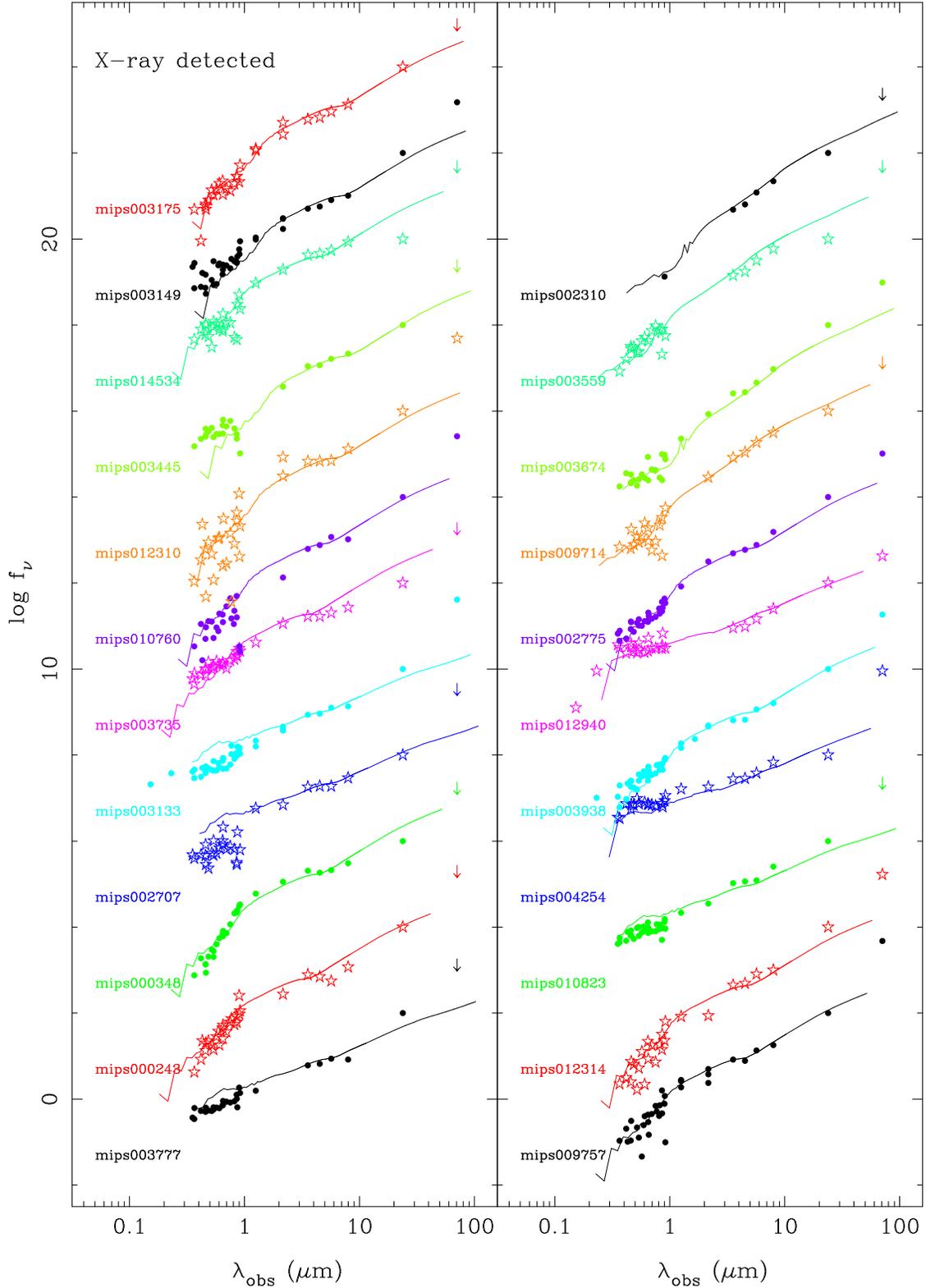}
\caption{Observed SEDs, arbitrarily offset for clarity, (filled dots and star 
symbols) of IR power-law galaxies detected 
in X-rays for which spectroscopic redshifts are not available. 
Each galaxy is shown with the closest matching SED template 
from those shown in Fig.~2 redshifted to the 
phot-z estimate. 
Note that this SED template is not the template used for computing the phot-z,
and is only used for source classification purposes (see text for more details).
For those galaxies not detected with MIPS at $70\,\mu$m 
the upper limits are set to the detection 
limit of $f_{70\,\mu{\rm m}} \simeq 3\,$mJy.}
\end{figure*}

\begin{figure*}
\vspace{-0.4cm}
\epsscale{0.9}
\plotone{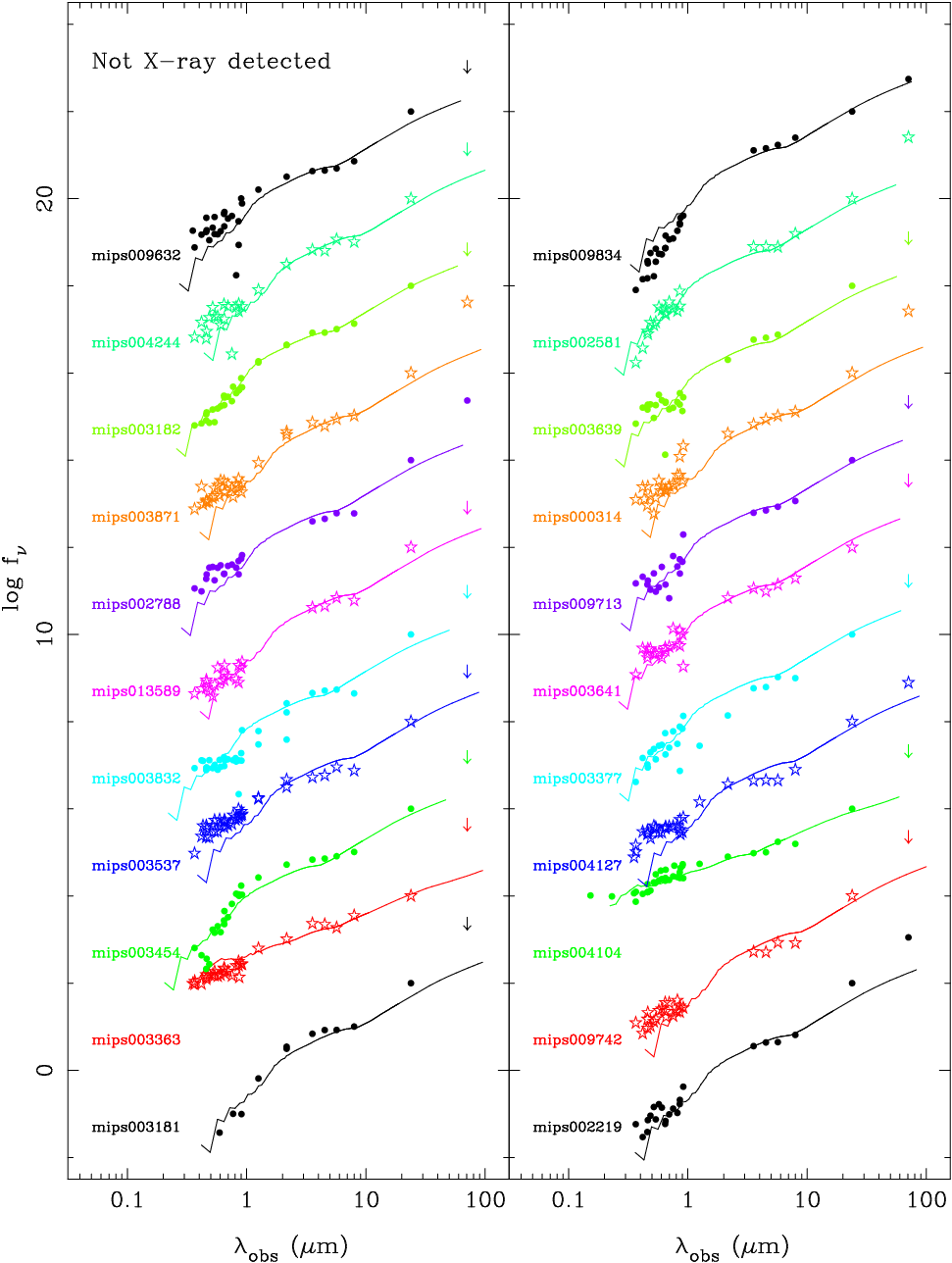}
\caption{Observed SEDs, arbitrarily offset for clarity,  (filled dots and star 
symbols) of IR power-law galaxies not detected 
in X-rays for which spectroscopic redshifts are not available. 
Each galaxy is shown with the closest matching SED template 
from those shown in Fig.~2 redshifted to the 
phot-z  estimate. Note that this SED template is not the 
template used for computing the phot-z, and is only used for 
source classification purposes (see text for more details). 
For those galaxies not detected with MIPS at $70\,\mu$m 
the upper limits are set to the detection 
limit of $f_{70\,\mu{\rm m}} \simeq 3\,$mJy.}
\vspace{0.4cm}
\end{figure*}

\begin{figure}
\setcounter{figure}{3}
\epsscale{0.5}
\vspace{0.4cm}
\plotone{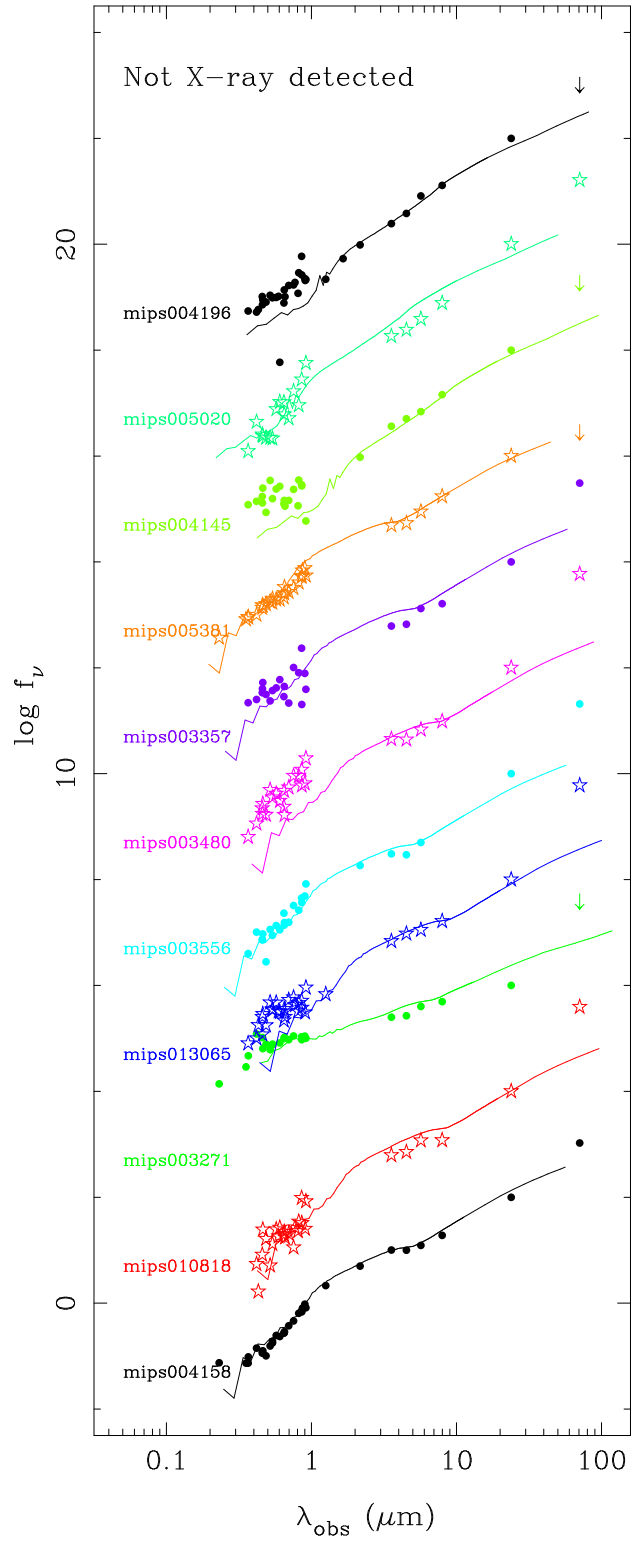}
\caption{Continued.}
\end{figure}

\subsection{X-ray detected galaxies with photometric redshifts}
For galaxies showing SEDs resembling the BLAGN SED category, virtually all of which are
known X-ray sources, we use mostly  phot-z estimates from COMBO-17 and
Zheng et al. (2004). Where redshifts are not available from
the literature, we determined redshifts with the code discussed
by P\'erez-Gonz\'alez et al. (2005). Briefly, the code uses
the observed optical to mid-IR SEDs and a full set of empirically
determined and theoretical templates (such as those in Fig.~2)
to derive phot-z's. As discussed by P\'erez-Gonz\'alez et al.
(2005), the main advantage of our phot-z code when used with
optical to mid-IR SEDs is the possibility of using the $1.6\,\mu$m
bump in addition to the optical stellar features routinely
employed by other phot-z codes. For galaxies with such a bump,
P\'erez-Gonz\'alez et al. (2005) demonstrate that the code gives
reliable redshifts our to z $\sim 3$  (compare also with the
results of Caputi et al. 2005, which confirm the general
results obtained with our code).

However, many of our sources do not have a prominent $1.6\,\mu$m 
bump. Rigby et al. (2005a) used the same code to estimate redshifts
for optically faint AGN, a substantially more difficult
problem than even the sources in our program. They found
that the values returned were in good agreement both with
redshifts from the literature, when available, and with those estimated
by eye. In particular, although they found a large range of
permissible redshifts for their sources, an equivalent range
in our estimates would not affect our conclusions.

Both P\'erez-Gonz\'alez et al. (2005) and Rigby et al. (2005a)
find difficulty in getting redshifts for purely power-law SEDs.
Fortunately, most such objects in our sample are in the BLAGN
group and redshifts are available from the literature based
on other redshift determined techniques. The great
majority of sources without redshifts depart from the power-law behavior
at short wavelengths, and hence have features or a general
shape that allows the phot-z code to return a value.
To test the reliability of these values, we have cross-correlated
our sample with the phot-z catalog of Zheng et al. (2004),
with COMBO-17 (Wolf et al. 2004), and for the objects with
spectroscopic redshifts, and we find good agreement for galaxies whose SEDs
resemble the NLAGN template. Since we use the
redshifts only for rough estimates of luminosities (\S6.1 and 7), precise
values are not required and we find that the phot-z's are
sufficiently accurate for our purposes.

Fig.~3 shows the observed SEDs for twenty-three out of the twenty-six 
galaxies in our sample detected in X-rays without spectroscopic redshifts;  
the three galaxies not shown here have  few points in their
SEDs and an estimate of their phot-z was not possible. 
We do not use the templates obtained
in this work for phot-z determination because of possible
interactions between the redshift and the type determination
for the SED. Our phot-z code instead uses actual SEDs of observed galaxies (see
P\'erez-Gonz\'alez et al. 2005 for a full description).
The SEDs of the majority of the IR power-law galaxies in Fig.~3 (see also
\S5.3, and Fig.~4) match one of the three SED categories defined in \S5.1. 
 For each galaxy in Fig.~3 we plot the closest matching
average template defined in this work,  not constrained to be the template used by our  phot-z
code, redshifted to the estimate of phot-z.  
The average SED templates are only used for source classification purposes,
and to estimate the rest-frame $12\,\mu$m flux density from the observed $24\,\mu$m value (see \S6.1).
The galaxies are shown in approximate order of increasing
IRAC slopes. The estimated phot-z's 
are between 0.7 and 2.8. 
Most of the galaxies without spectroscopic redshifts show
SEDs similar to the NLAGN template as they tend to be optically fainter than
those with spectroscopic redshifts. A few galaxies with very 
steep IRAC slopes resemble the local ULIRG Mrk~231.

\subsection{Galaxies not detected in X-rays}
Among the galaxies not detected in X-rays, we have only found one galaxy 
in the VVDS spectroscopic redshift sample (Le F\`evre et al. 2004). 
For the rest of the galaxies not detected 
in X-rays we have estimated phot-z's  and analyzed the SEDs 
using the same technique as in \S5.2. 
Fig.~4 shows the SEDs (in order of increasing IRAC slope) for 
the 32 IR power-law galaxies not detected in X-rays for which we could
estimate 
phot-z's plus the galaxy with the VVDS spectroscopic redshift; 
the other 10 sources have incomplete SEDs and a reliable estimate 
of the phot-z was not possible.

The majority of these sources have SEDs
similar to the NLAGN template; a few show very steep SEDs similar to that
of Mrk~231. Only three galaxies show an SED similar to that of a BLAGN. One
of these sources (mips003271) is classified by COMBO-17 as a QSO based on the
presence of broad lines with a phot-z of 
$z_{\rm COMBO} = 1.63$. This galaxy is located at a large X-ray off-axis angle in
the {\it Chandra} images (see
Table~2) where the X-ray sensitivity is less, and the upper limit to its X-ray
luminosity is consistent with the presence of an AGN. 
The second source (mips004104) appears to be a relatively nearby galaxy 
($z_{\rm COMBO} = 0.31$), but its SED is similar to that of a
broad line AGN. The third source (mips003363) has a COMBO-17 phot-z of 
$z_{\rm  COMBO} = 1.2$, and is also located at a large off-axis angle. 
The phot-z's of the IR power-law
galaxies not detected in X-rays are between 0.3 and 3  (see insert of 
Fig.~5).

\begin{figure}
\epsscale{0.8}
\plotone{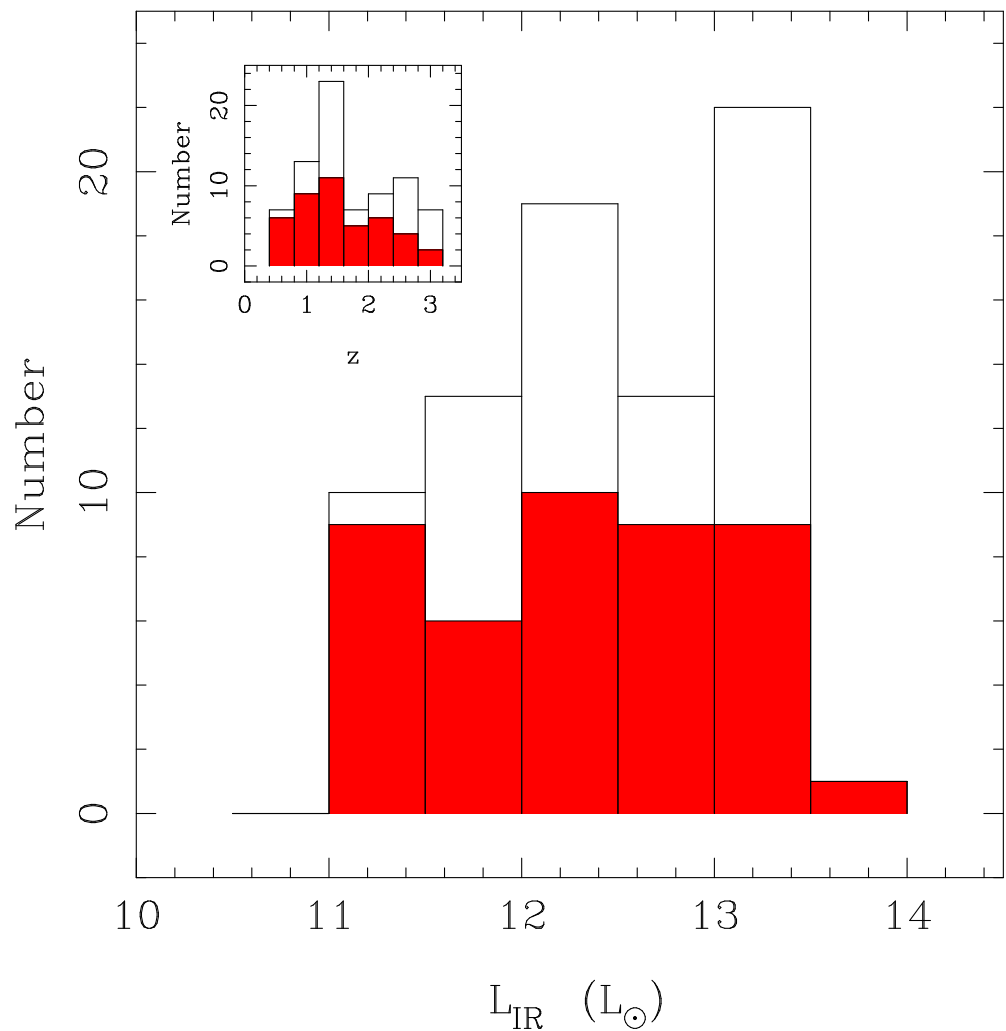}
\caption{{\it Main panel:} Distribution of the 
rest-frame IR $8-1000\,\mu$m luminosity 
for the entire sample of IR power-law galaxies with redshifts (empty
histogram) and  those detected in X-rays (filled histogram). {\it Insert:} 
Distribution of redshifts (both spectroscopic and photometric) of the 
IR power-law galaxies (empty histogram), and the galaxies detected in 
X-rays (filled histogram).}
\end{figure} 

\subsection{Summary of properties}
The SEDs of IR power-law galaxies detected
in X-rays are approximately equally divided between the  BLAGN SED category (40\%) and
NLAGN+ULIRG SED category (60\%). As the division between the NLAGN and the ULIRG SED category is rather
arbitrary, in what follows we discuss both categories jointly.
The galaxies not detected in X-rays are mostly in the
NLAGN+ULIRG SED category, consistent with the fact that the majority of these
galaxies are optically faint --- most have 
COMBO-17 $R$-band magnitudes
$R>24\,$mag (see Appendix) --- and are possibly more obscured than the 
X-ray detected ones.  The distribution of $24\,\mu$m flux densities near the
completeness limit, on the other hand,  is similar for
galaxies detected and not detected in X-rays (see Fig.~1). 
Both the galaxies detected in X-rays and those not detected in
X-rays show a similar fraction of $70\,\mu$m
detections ($\simeq 40\%$) down to an approximate limit of 
$f_{70\,\mu{\rm m}} \simeq 3\,$mJy (see Table~1).

In terms of the redshift distribution (see insert of Fig.~5) the majority of
the IR power-law galaxies (both those detected in X-rays and those 
not detected in X-rays) appear to be at $z>1$. This is in contrast with 
the two observed peaks in the spectroscopic redshift distribution 
of the X-ray sources in the CDFS ($z=0.674$ and 
$z=0.734$, Szokoly et al. 2004) and  the spectroscopic 
redshift grouping of galaxies with $I_{\rm AB} \le 24$ observed by VVDS
($z=0.73$, Le F\`evre et al. 2004).
There seems to be a slightly larger proportion of  IR power-law
galaxies at $z>2$ not detected in X-rays (15 out of 33) than detected in
X-rays (12 out of 46), but this result needs to be confirmed
spectroscopically. 
Most of the IR power-law galaxies at $z<1$ are detected in X-rays.

\begin{figure*}
\epsscale{1.}
\plotone{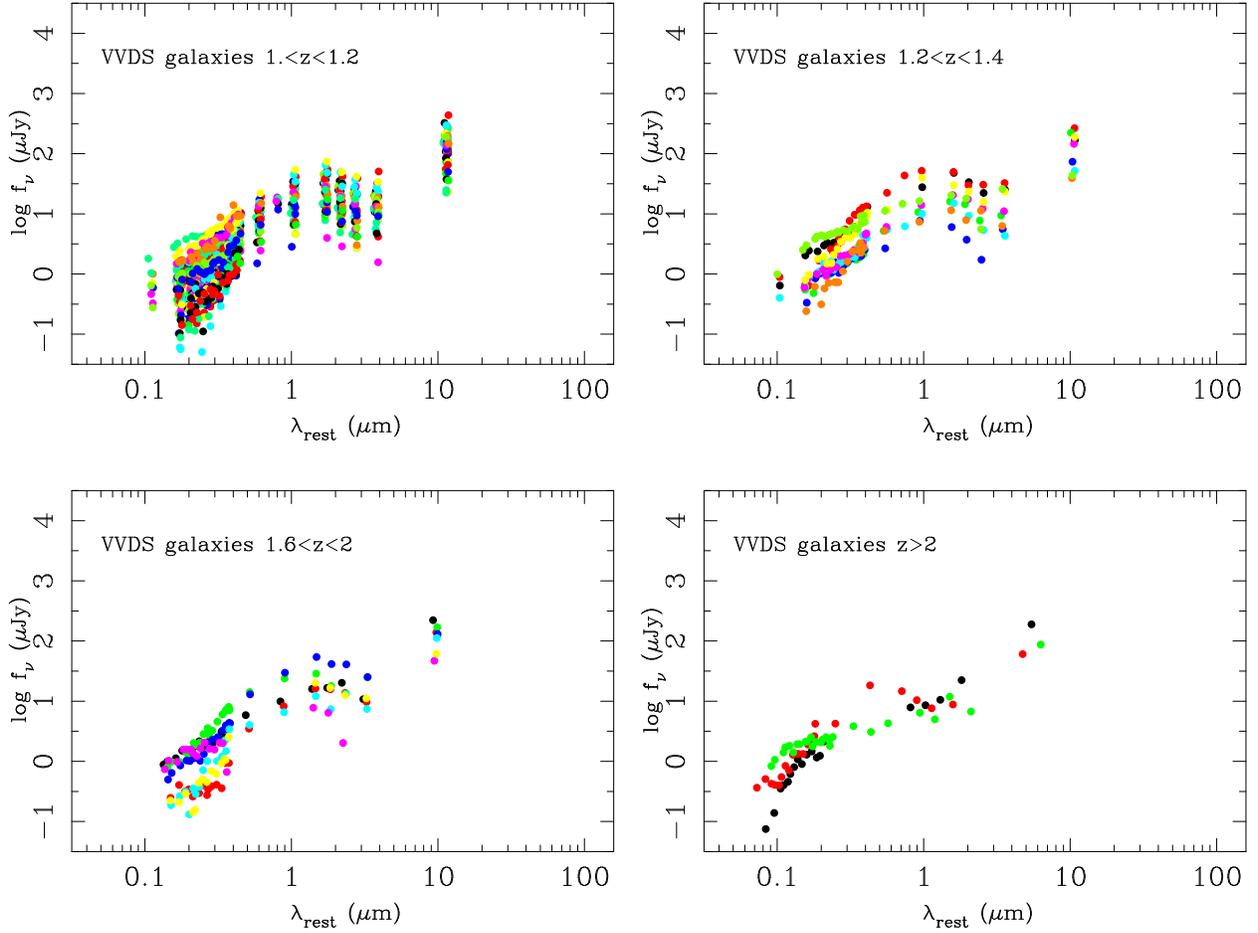}
\caption{SEDs of VVDS galaxies ($I_{\rm AB} \le 24$) 
in the CDFS at $z>1$ included in the sample of
 predominantly star-forming galaxies selected at 
$24\,\mu$m by P\'erez-Gonz\'alez et al. (2005). The
  galaxies are plotted in different redshift intervals. At $1<z<2$ the
  $1.6\,\mu$m stellar bump is very prominent, unlike the case of the IR power-law
  galaxies (see Figs.~2, 3, and 4). }
\end{figure*} 

\section{Infrared properties}
\subsection{Rest-frame $12\,\mu$m and Total IR Luminosities}

While the total IR ($8-1000\,\mu$m) luminosity cannot be measured directly for
most of the IR power-law galaxies in our sample, a number of works 
(e.g., Rush, Malkan, \& Spinoglio 1993; Spinoglio et
al. 1995; 
Takeuchi et al. 2005; P\'erez-Gonz\'alez et 
al. 2005) have shown that the $12\,\mu$m 
luminosity is a good proxy for it for AGNs, LIRGs, and ULIRGs. 

We have  computed the $12\,\mu$m to IR
luminosity conversion
factors from warm ($f_{25\mu{\rm m}}/
f_{60\mu{\rm m}} >0.2$) and cool ($f_{25\mu{\rm m}}/
f_{60\mu{\rm m}} <0.2$, see Sanders \& Mirabel 1996) ULIRGs in the local
Universe (those discussed in 
Appendix~A.3) to be used for IR power-law galaxies in the BLAGN and NLAGN+ULIRG SED
classes, respectively. Although our procedure to compute IR luminosities 
is very similar to that used by P\'erez-Gonz\'alez et al. (2005),
we have taken special care to use $12\,\mu$m to IR luminosity ratios specific to the
class of galaxies in study. In particular, galaxies whose SEDs resemble those
of optical QSOs show $12\,\mu$m to IR luminosity ratios significantly 
lower than the typical
values of cool ULIRGs and some warm ULIRGs (e.g., Mrk~231). 
The $12\,\mu$m rest-frame luminosities 
are derived from the observed $24\,\mu$m values 
using the spectral indices of power laws fitted in the $\lambda_{\rm
  rest}=10-30\,\mu$m range 
of the average
templates defined in \S5.1.
The IR luminosities are only
computed for those galaxies with spectroscopic redshifts and 
reliable estimates of the phot-z.

All the IR power-law galaxies are highly luminous.  About 30\% are in the
hyperluminous class ($L_{\rm IR} > 10^{13}\,{\rm L}_\odot$; 
Kleinmann et al. 1989, and see also Rowan-Robinson 2000), 41\% are ULIRGs 
($L_{\rm IR} = 10^{12}-10^{13}\,{\rm L}_\odot$), and all but one of the rest
are LIRGs ($L_{\rm IR} = 10^{11}-
10^{12}\,{\rm L}_\odot$).  Fig.~5 shows the distribution of IR luminosities.
At the lower 
IR luminosity end ($L_{\rm IR} < 10^{12}\,{\rm L}_\odot$) a large fraction
are detected in X-rays and are mostly in the BLAGN SED category.
The fraction of X-ray detected
galaxies in the sample of IR power-law galaxies 
remains approximately constant (50\%) at $L_{\rm IR} > 10^{12}\,{\rm  L}_\odot$.  

As we saw in \S5, the majority of the IR power-law galaxies
have SEDs similar to the average SED templates shown in Fig.~2. If 
we extrapolated these templates, the IR power-law galaxies 
would satisfy the criterion for the warm ULIRG 
class, so it is likely that a large fraction of
 the IR power-law galaxies in
the ULIRG category, together with galaxies detected in X-rays, contain an AGN.

\subsection{IR Power-Law galaxies Compared With High-$z$ Star-forming Galaxies}
P\'erez-Gonz\'alez et al. (2005) and LeFloc'h et al. (2005) 
have studied the evolution of the IR
luminosity of $24\,\mu$m-selected galaxies in the CDFS and the Hubble Deep
Field North (HDFN) out to a redshifts of $z=1$ and $z=3$, respectively. They find that 
ULIRGs dominate the galaxy population at $z>2$, and make a 
significant contribution ($\simeq 50\%$) 
in the $1<z<2$ redshift interval, whereas at $z<1$ they appear to be extremely
rare. They excluded galaxies with SEDs
increasing monotonically from the optical to the IR, that is, galaxies similar
to those in our sample, because they were interested in purely star-forming 
galaxies. 

The IR  power-law galaxies 
constitute a small fraction, approximately $10-15$\%, of this ULIRG 
population at $z>1$. 
This fraction is  
similar to that of warm SED ULIRGs among SCUBA/VLA selected high-$z$ 
sources in the Lockman Hole (Egami et al. 2004), the fraction of AGN
identified by means of the shape of the SED in 
high-$z$ submillimeter galaxies (Ivison et al. 2004), and the percentage of warm 
ULIRGs in the local Universe. 

Egami et al. (2004) showed that the 
SEDs of their SCUBA/VLA selected high-$z$ galaxies divided into cool and warm ULIRGs, 
with the former showing a significant stellar bump at $1.6\,\mu$m. 
To further investigate the nature of the IR power-law ULIRGs, we will compare
them with high-$z$ star-forming LIRGs and ULIRGs in P\'erez-Gonz\'alez et al. (2005). We
have used all the IR ($24\,\mu$m) selected galaxies  at $z>1$ 
with spectroscopic redshifts from the VVDS without evidence for the presence of an
AGN (i.e., galaxies without an X-ray detection and/or a QSO spectroscopic identification). 
Their SEDs are shown in Fig.~6 in
different redshift bins. 

The SEDs of star-forming galaxies selected at $24\,\mu$m and
identified by the VVDS in 
the $1<z<2$ redshift range show a very pronounced $1.6\,\mu$m stellar bump
(there are only three VVDS galaxies at $z > 2$ and we cannot assess the nature of their SEDs),
unlike the IR power-law galaxies (see Figs.~2, 3, and 4). 
This result tends to confirm the arguments in Appendix~A based on local
LIRGs and ULIRGs that the power-law behavior is not characteristic of star formation, but
it indicates the presence of an AGN. (We cannot 
exclude intense star formation in the IR power-law galaxies, or even the possibility that 
star formation could dominate the bolometric luminosity of these galaxies.) 
One caution is that these VVDS objects are optically bright
($I_{\rm AB} \le 24$) and therefore are probably biased toward very massive galaxies.
That is, this spectroscopic sample
probes the high end of the mass function at a given redshift,
revealing galaxies that contain a significant population of evolved stars which
might dominate the total mass of the galaxy and make the $1.6\,\mu$m bump
very prominent.

\begin{figure}
\epsscale{0.8}
\plotone{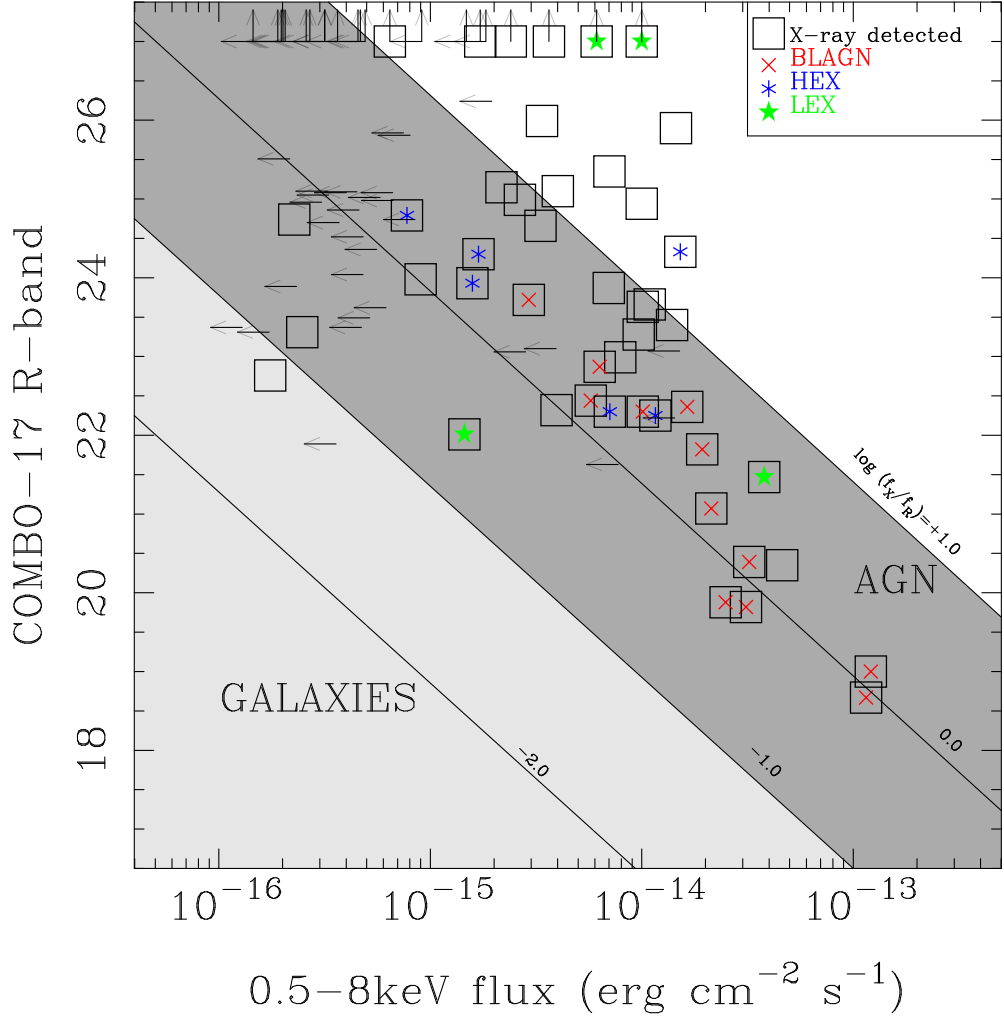}
\caption{COMBO-17 $R$-band magnitudes (Wolf et al. 2004) vs. the 
{\it Chandra} full-band X-ray fluxes for 
X-ray detections and non-detections in the IR power-law sample. When
available, 
we show the optical spectroscopic 
classification from Szokoly et al. (2004). Those galaxies
in our sample not detected by COMBO-17 are displayed at an upper limit of
$R=27\,$mag. The lightly shaded area is the approximate location of
``normal'' (star-forming) galaxies, and the dark shaded area is the location
of AGN, adapted from Bauer et al.  (2004). }
\end{figure}

\begin{figure}
\epsscale{0.8}
\plotone{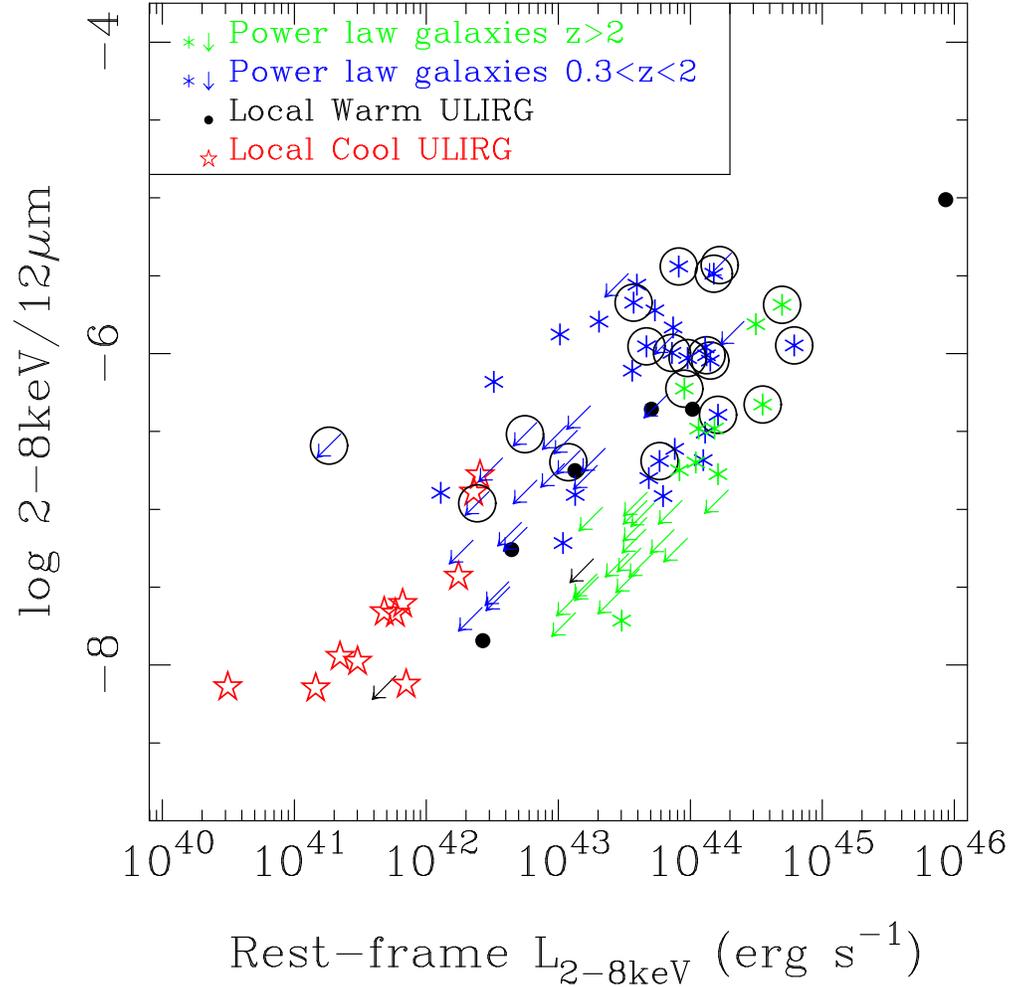}
\caption{Rest-frame $2-8\,$keV luminosity  
vs. the rest-frame hard X-ray to $12\,\mu$m flux ratio for the IR power-law
SED galaxies (asterisks) compared with local examples of warm 
(AGN dominated, filled dots) and  cool ULIRGs (with and without 
AGN in their nuclei, open star symbols). For the warm ULIRG sample we use
all the galaxies in the original sample of Sanders et al. (1988) with
detections in all four IRAS bands. We note that a few of those galaxies are
not strictly ULIRGs, that is, they have $L_{\rm IR} < 10^{12}\,{\rm L}_\odot$, but they have
been optically identified as AGN. The X-ray data for the local 
Universe ULIRGs are from Reeves \& Turner (2000), Boller et al.
(2002), Ptak et al. (2003), Franceschini et al. (2003), and 
Risaliti et al. (2000, and references therein). The IR power-law 
galaxies have been divided into two redshift intervals.
The open circles denote those IR power-law galaxies in the BLAGN SED category.}
\end{figure} 

\begin{figure*}
\epsscale{1.}
 
\plotone{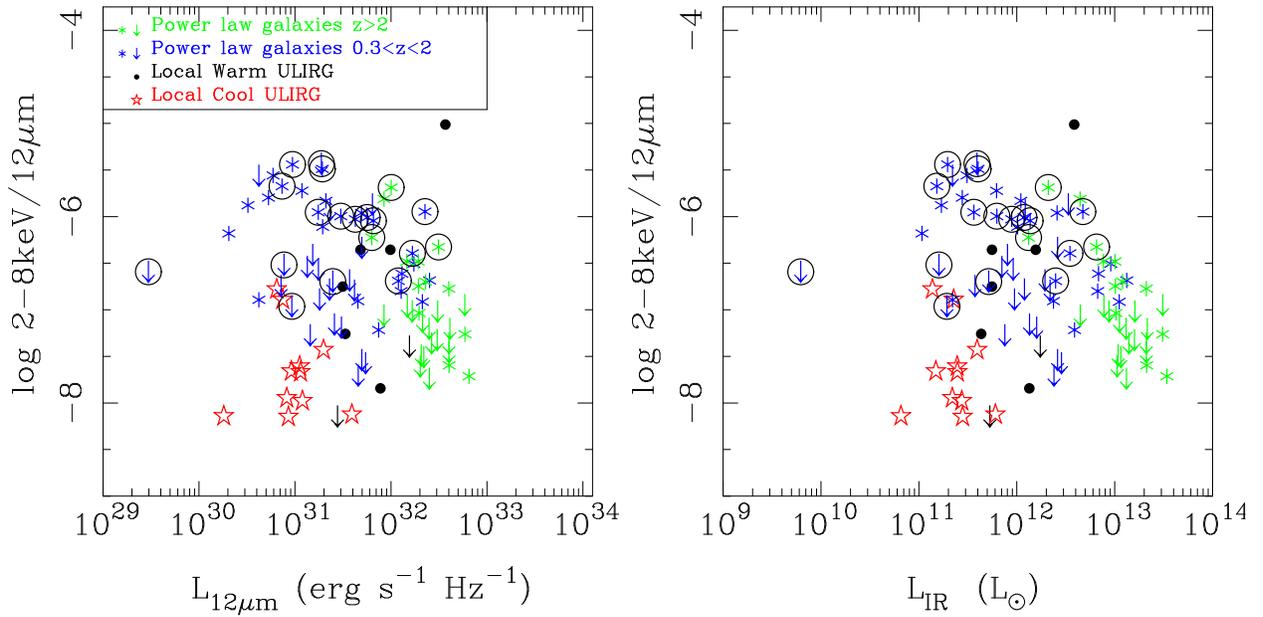}
\caption{Rest-frame 12$\,\mu$m luminosity (left panel) 
and total IR $8-1000\,\mu$m luminosity (right panel), both
computed as described in \S6.1, 
vs. the rest-frame hard X-ray to mid-IR flux ratio for the power-law
SED galaxies (asterisks) compared with local examples of warm 
(AGN dominated, filled dots) and  cool ULIRGs (open star symbols).
The open circles denote those IR power-law galaxies in the BLAGN-SED category.}
\end{figure*}

\section{X-ray properties}
\subsection{X-ray vs optical} 
Bauer et al. (2004, and references therein) have demonstrated that X-ray-to-optical 
flux ratios can be useful for distinguishing between AGN and
star-forming galaxies for sources detected in deep ($1-2\,$Ms) X-ray
exposures. In particular, one of the criteria these authors used to classify
sources as AGN is $\log f_{\rm 0.5-8keV} /f_R >-1$, whereas 
star-forming and normal galaxies have $\log f_{\rm 0.5-8keV} /f_R <-1$. 
We have used this approach to probe the nature of the IR power-law galaxies. Fig.~7 compares
the COMBO-17 $R$-band magnitudes and the observed 
$0.5-8\,$keV fluxes for the galaxies
in our sample. We also show the regions occupied by AGN and star-forming galaxies, adapted
from Bauer et al. (2004). All but one of the IR power-law galaxies detected in
X-rays 
are located in the area occupied by AGN ($\log f_{\rm 0.5-8keV} /f_R
>-1$), and their place in the diagram is consistent with these galaxies having
X-ray luminosities  $L_{\rm 0.5-8keV} > 10^{42.5}\,{\rm erg\,s}^{-1}$ (see
Bauer et al. 2004, and 
below). None of the IR power-law galaxies detected in X-rays would be 
classified as a star-forming galaxy based on its X-ray luminosity (i.e., 
$L_{\rm 0.5-8keV} < 10^{42}\,{\rm erg\,s}^{-1}$). 

Most of the IR power-law galaxies not detected in X-rays tend to occupy a
similar region as the X-ray detected galaxies (but as X-ray upper limits) in 
the $R$-band vs. $0.5-8\,$keV diagram. According
to Bauer et al. (2004) the location of these galaxies in Fig.~7 is consistent
with them having upper limits to their X-ray luminosities above
$L_{\rm 0.5-8keV} \simeq 10^{41.5}\,{\rm erg\,s}^{-1}$ (see below). By this
test, they are either very obscured in the X-rays or they could be dominated
by star-formation. The lack of a $1.6\,\mu$m evolved stellar bump argues for
the former explanation.

For X-ray sources in the CDFS, Szokoly et al. (2004) find that most X-ray 
type 2 QSOs (that is, galaxies absorbed in X-rays) have $R>24\,$mag, whereas
type 1 QSOs tend to be optically bright (see also our Fig.~7 since there is a
good correspondence between BLAGN and X-ray type 1 QSOs). If the IR power-law
galaxies not detected in X-rays in our sample
contain an AGN, their optical magnitudes make them candidates to be X-ray type
2 (absorbed) AGN, also consistent with the fact that they 
show SEDs similar to the NLAGN and ULIRG categories.

\subsection{X-ray vs. IR}

As shown in \S6, a large fraction of the galaxies in our sample are 
ULIRGs. Local Universe ULIRGs are known
to be intrinsically faint hard X-ray emitters --- typically 
$L_{\rm 2-10keV} = 10^{41} - 10^{43}\,{\rm erg \, s}^{-1}$ (see Risaliti et
al. 2000; Franceschini et al. 2003; Ptak et al. 2003, and also Fig.~8). Warm ULIRGs
and those containing AGN tend to have higher X-ray luminosities
(Fig.~8). The origin of the hard X-ray continuum of ULIRGs is still 
uncertain, but for some cool ULIRGs the extended nature of their soft X-ray
emission and the shape of the X-ray spectra seem to indicate that the X-rays 
are  produced by star formation
(Franceschini et al. 2003, Ptak et al. 2003). Risaliti et al. (2000) 
found that there is a significant
fraction of LIRGs and ULIRGs optically classified as AGN
that do not show any evidence of X-ray activity even in the {\it BeppoSAX} 
$20-200\,$keV band, suggesting that these sources are completely Compton thick
($N_{\rm H} > 10^{25}\,{\rm cm}^{-2}$).

To further understand the nature of the IR power-law galaxies in
the CDFS, their X-ray and IR properties are compared with their 
local Universe analogs in Figs.~8 and 9. 
For the IR power-law galaxies we calculate the rest-frame $2-8\,$keV X-ray
luminosities (not corrected for absorption) 
assuming an intrinsic  photon index of $\Gamma = 1.8$ for which the k-corrections
at low redshift ($z<2$) are relatively small
(Barger et al. 2005), and using equation~1 from Alexander 
et al. (2003b). Fig.~8 shows
the rest-frame $2-8\,$keV to $12\,\mu$m flux ratio vs. the rest-frame 
$2-8\,$keV luminosity. Fig.~9 shows the same ratio vs. the rest-frame
$12\,\mu$m and IR luminosities for the IR power-law galaxies.

The majority of the galaxies in our sample in the BLAGN SED class show 
hard X-ray luminosities $L_{\rm 2-8keV} \gtrsim  10^{43}-10^{44}\,{\rm erg \, s}^{-1}$ (see 
Fig.~8), as 
demonstrated among others by Zheng et al. (2004), Szokoly et al. (2004), and 
Barger et al. (2005), based on spectroscopic classifications. 
Their hard X-ray to mid-IR ratios are similar 
to those of X-ray selected galaxies in 
the local Universe (e.g., the Piccinotti et al. 
1982 sample).

The IR power-law
galaxies in the NLAGN+ULIRG SED category show a broad range of hard X-ray to
mid-IR ratios (also observed in general for X-ray sources with $24\,\mu$m
counterparts, see Rigby et al. 2004), although in general below those of the BLAGN SED
category. They show hard X-ray luminosities in the range $L_{\rm 2-8keV} = 
10^{42}-10^{44}\,{\rm erg \, s}^{-1}$.
 Both in terms of the hard X-ray luminosities and 
${\rm 2-8keV}/12\mu{\rm m}$ ratios, the NLAGN SED galaxies in our sample 
are similar to the local
Universe warm ULIRGs. Local cool ULIRGs, on the other hand, tend to have hard X-ray
luminosities (not corrected for absorption) $L_{\rm 2-8keV} 
< 2 \times 10^{42}\,{\rm erg \, s}^{-1}$. 
We find
that the IR power-law galaxies detected in X-rays 
with values of the photon index typical of highly absorbed or Compton
thick X-ray
emission ($\Gamma < 1$, Bauer et al. 2004) tend to have 
$\log  ({\rm 2-8keV}/12\mu{\rm m}) \lesssim -6.5$,
which could explain the smaller hard X-ray to mid-IR
flux ratios if the rest-frame hard X-rays are heavily absorbed. 

The IR power-law galaxies not detected in X-rays tend to have upper limits 
to the $0.5-8\,$keV luminosities and the ${\rm 2-8keV}/12\mu{\rm m}$ 
ratios in the observed range of the NLAGN SED galaxies
detected in X-rays. These properties are also similar to those of some of the
warm ULIRGs in the local Universe.

In terms of the rest-frame $12\,\mu$m luminosities, the IR power-law galaxies in
the BLAGN SED category span a larger range than in hard X-ray luminosities
(Fig.~8 and 9).
The high-$z$ IR power-law galaxies in the NLAGN+ULIRG SED class tend to be amongst 
the most luminous objects at $12\,\mu$m and in the IR. 
Since there is a larger overlap in
$12\,\mu$m luminosities between the BLAGN and NLAGN+ULIRG SED categories than
in hard X-ray luminosities, it
is likely that the smaller ${\rm 2-8keV}/12\mu{\rm m}$ ratios in the
latter class are due to absorbed X-ray emission, as also suggested by the
fitted photon indices, unless BLAGN SED galaxies are intrinsically more
X-ray luminous than the NLAGN+ULIRG SED galaxies.

\subsection{X-ray stacking of undetected galaxies}
We stacked the X-ray data of the 17 individually undetected galaxies 
at off-axis angles of less than $9\arcmin$ and lying  further away than 3
times  the 90\% encircled energy radii of other 
X-ray sources in the field, following 
the procedure of Immler et al. (2005) and Worsley et al. (2005). This 
yielded a significant detection in the soft band ($3.9\sigma$) and a tentative
detection in the hard band ($2.2\sigma$). The corresponding average
soft-band and hard-band fluxes are
$\approx6.5\times10^{-18}$~erg~cm$^{-2}$~s$^{-1}$ and
$\approx2.7\times10^{-17}$~erg~cm$^{-2}$~s$^{-1}$, calculated assuming
$\Gamma=1.4$; assuming a non detection in the hard band, the $3\sigma$
upper limit is $<5.6\times10^{-17}$~erg~cm$^{-2}$~s$^{-1}$. The average
X-ray spectral slope is $\Gamma\approx0.9$ (hard-band detected) or
$\Gamma>0.3$ (hard-band undetected). For $z=2$ the observed soft and hard-band
luminosities of the detections would be $2\times 10^{41}\,{\rm erg \ s}^{-1}$ and 
$8\times 10^{41}\,{\rm erg \ s}^{-1}$, respectively.
This is consistent with obscured AGN.

\section{Comparison with Predictions for Obscured Infrared AGN}

Ueda et al. (2003) and Treister et al. (2004) have put forward models to reconcile the
observed redshift distribution of AGN and that predicted
by the $2-10\,$keV X-ray background synthesis models. Ueda et al. (2003)
derive a ratio of about 1:1 for hard X-ray detections of obscured 
($N_{\rm H} > 10^{22}\,{\rm cm}^{-2}$) to unobscured AGN. 
Treister et al. (2004) specifically modeled the 
Great Observatories Origins Deep Survey (GOODS) CDFS and HDFN
 fields. 
They also predict that a faint $24\,\mu$m-selected sample will have 
about a 2:1 to 3:1 ratio of obscured to unobscured AGN. 
In the deep {\it Chandra} fields the spectroscopically observed 
fraction of obscured to unobscured AGN is 2 to 1 at $z \simeq 0.7$ 
(Barger et al. 2003).  

We now compare these predictions with our results on IR power-law galaxies and
X-ray detected AGN. The FOV of our study is approximately
$0.1\deg^2$, although the X-ray sensitivity is not uniform across the field. 
Of the CDFS galaxies detected in the hard X-ray 
band ($> 2 \times 10^{-16}\,{\rm erg\,cm}\,^{-2}\,
{\rm s}^{-1}$, to ensure that they are AGN),  
116 are detected at $24\,\mu$m down to approximately 
80$\mu$Jy, where the 24$\mu$m
counts are $\sim$ 80\% complete. Treister et al.
(2004) infer that among those CDFS AGN with spectroscopic 
redshifts about 55\% have $N_{\rm H} > 10^{22}\,{\rm cm}^{-2}$
(see also Rigby et al. 2004). Assuming
that this fraction remains the same for those X-ray sources detected at 
$24\,\mu$m, then we would expect approximately 
52/0.8 = 65 unobscured and 64/0.8 = 80 obscured AGN among the hard X-ray 
sources detected at $24\,\mu$m. For a FOV of $0.1\deg^2$  
Treister et al. (2004) models would predict $\sim$ 88 unobscured 
sources (type 1) and $\sim$ 180 - 250 obscured AGN 
with $f_{24\mu {\rm m}} \gtrsim 80\,\mu$Jy. The smaller number of observed
unobscured sources is probably due to the fact that we are using 
the entire X-ray FOV, whereas the Treister et al. (2004) predictions
for the GOODS field are for the area with the most
sensitive X-ray observations (see below). 

How many of these obscured galaxies can be accounted for? 
In our sample of IR power-law galaxies, we have $54/0.8=68$ galaxies not detected 
in hard X-rays (see Table~2). As we showed in \S6 most of these galaxies 
have IR luminosities above $2 \times 10^{12}\,{\rm L}_\odot$. 
Moreover, the IR power law galaxies not detected
in X-rays have upper limits to the 
rest-frame ${\rm 2-8keV}/12\mu{\rm m}$ flux ratios that are typical of absorbed AGN. 
Adding the power-law IR galaxies not detected in hard X-rays and the 
predicted number of  obscured hard X-ray sources detected at $24\,\mu$m, 
we would have approximately 148 obscured AGN, and thus a ratio of
obscured to unobscured $24\,\mu$m detected AGN of 2:1. 

If we restrict ourselves to the most sensitive
X-ray area (off-axis angles $\le 8\arcmin$, area $\simeq 0.05\deg^2$), 
following the same reasoning as above, we find 90 obscured 
AGN and 47 unobscured
AGN at $24\,\mu$m. Treister et al. (2004) predict an AGN number count at $24\,\mu$m of
approximately $100-130$ for a field this size, consistent with our data.  
The observed obscured to unobscured ratio of $24\,\mu$m detected AGN remains
2:1 for the most sensitive X-ray area.

Thus, although our 
results are consistent with the lower obscured to unobscured ratio of 
Treister et al. (2004), we have not identified all of the predicted
fraction of obscured AGN. 
However, it is likely that the {\it Spitzer} 
data will eventually reveal a significant number of additional obscured AGN. 
In the 1\,Jy sample of ULIRGs, Veilleux et al. 
(1999) find that approximately 30\% of galaxies are optically classified as 
Seyfert. If the fraction in the local Universe is similar to that at higher redshifts, it is
possible that we are still missing some 50 ULIRGs (with no X-ray detections) 
from our sample of IR power-law galaxies 
containing an AGN at $z>1$. Specifically, many galaxies may be analogous to
the cool ULIRGs in the local Universe that host an AGN but have SEDs 
not dominated by the AGN emission  (e.g., Mrk~273, UGC~5101), and are 
not bright in X-rays (see \S7.2). A significant number of 
additional obscured AGN can be identified by comparing radio and IR
measurements. For example, Donley et al. (2005) find 16 radio-bright 
AGN in the HDFN that are above
our threshold for 24$\mu$m detection but 
below the threshold for the X-ray catalogs of the region. 

\section{Conclusion}

The presence of power-law emission in the optical and/or IR is usually
taken to indicate an AGN. In this paper, we investigated the
nature of the IR power-law galaxies in the CDFS. We emphasized
probing whether those not detected in X-rays are part of a
population of obscured AGN.

We used  {\it Spitzer}/MIPS $24\mu$m  observations 
($f_{24\mu {\rm m}} \gtrsim 80\,\mu$Jy)
in the CDFS to select a sample of 92 galaxies 
showing power-law-like emission in the IRAC bands ($f_\nu \propto \nu^\alpha$, with 
$\alpha < -0.5$). 
Approximately 53\% (49 galaxies) of these IR power-law
galaxies are detected in at least one of the {\it Chandra} X-ray bands, and they
comprise approximately 30\% of the hard X-ray sources with $24\,\mu$m
counterparts in the CDFS. For those IR power-law galaxies not detected in
X-rays we have derived upper limits to the {\it Chandra} X-ray band fluxes that are
consistent with their harboring AGN.

The main properties of our sample of IR power-law galaxies are as follows:

\begin{itemize}


\item
{\it Spectroscopic and photometric redshifts}. The redshifts of the IR power-law galaxies 
are between 0.6 and 3, although
most of the galaxies in the sample lie at $z>1$, in contrast with the
observed spectroscopic grouping at 
$z\simeq 0.7$ for X-ray sources and galaxies with $I_{\rm AB} \le
24$ in the CDFS. 

\item
{\it Shape of the SED}.  The
IR power-law galaxies detected in X-rays are almost equally divided between BLAGN
(40\%) and NLAGN+ULIRG (60\%) classes, whereas the majority of the galaxies
not detected in X-rays have SEDs in the NLAGN+ULIRG class as they tend to be
optically fainter, and possibly more obscured. The steepest IRAC SEDs correspond to
$\nu$$^{-2.8}$, which is also a limit found for optically-faint X-ray AGN and
for a complete sample of flat spectrum radio sources. This slope evidently defines
limiting behavior for the optical-IR SEDs of AGN.

\item
{\it IR Luminosities}. 
Approximately 40\% of the sample of IR power-law galaxies are ULIRGs, and 
a further 30\% are in the hyperluminous IR galaxy class. At the lower
IR luminosity end ($L_{\rm IR} = 10^{11}-10^{12}\,{\rm L}_\odot$) the majority of
galaxies are detected in X-rays, particularly for those in the BLAGN SED category.
At the upper end of the IR luminosity 
distribution the
fraction of X-ray detected galaxies is approximately 50\% of the sample of IR
power law galaxies, although the majority of the IR power-law galaxies not detected in
X-rays have IR luminosities above $10^{12}\,{\rm L}_\odot$.

\item
{\it IR power-law galaxies vs. IR selected star-forming galaxies at high z.}
The ULIRGs in the IR power-law galaxy 
sample represent a small fraction, $10-15\%$, of the high-$z$ ULIRG population
of predominantly star-forming galaxies
detected with {\it Spitzer} (P\'erez-Gonz\'alez et al. 2005).
The SEDs of IR  ($24\,\mu$m selected) star-forming galaxies identified 
by the VVDS ($I_{\rm AB} \le 24$) in the $1<z_{\rm sp}<2$ redshift
interval 
are  dominated by stellar emission unlike the majority of the IR power-law galaxies.
\item
{\it X-ray Properties of X-ray detected IR power-law galaxies.}
All the X-ray detected galaxies in our sample have rest-frame $2-8\,$keV  
luminosities above 
$10^{42}\,{\rm erg \,s}^{-1}$.
Galaxies in the BLAGN SED category have rest-frame
hard X-ray luminosities $L_{\rm 2-8keV} \gtrsim  10^{43}-
10^{44}\,{\rm erg \, s}^{-1}$, and their X-ray to mid-IR flux ratios are  
similar to 
those of local X-ray selected galaxies. X-ray galaxies in the
NLAGN+ULIRG SED class show in general lower X-ray luminosities
($L_{\rm 2-8keV} = 
10^{42}-10^{44}\,{\rm erg \, s}^{-1}$) and 
X-ray to mid-IR ratios lower than those of the BLAGN SED galaxies, 
as well as X-ray photon indices
indicative of absorbed X-ray emission.  

\item
{\it X-ray Properties of IR power-law galaxies not detected in X-rays.}
For these objects, the upper limits to the rest-frame $2-8\,$keV  
luminosities ($L_{\rm 2-8keV} < 10^{42}-10^{44}\,{\rm erg \,s}^{-1}$) and 
to the hard X-ray to mid-IR flux ratios are similar to 
those of the X-ray detected galaxies in the NLAGN+ULIRG SED class, and
consistent with those of some of the
local Universe warm 
ULIRGs known to contain an AGN.
\end{itemize}

The upper limits to the  X-ray luminosities and hard X-ray to mid-IR flux
ratios, as well as the shape of the SEDs, the optical magnitudes, 
and the IR luminosities of the IR power-law galaxies not detected in X-rays make them 
good candidates to contain IR-bright X-ray-absorbed AGN. 
Adding together the IR power-law
galaxies not detected in hard X-rays and the fraction of absorbed 
($N_{\rm H} > 10^{22}\,{\rm cm}^{-2}$) X-ray
galaxies detected at $24\,\mu$m estimated by Treister et al. (2004), we find 
a ratio of obscured to unobscured AGN
detected at $24\,\mu$m of about 2:1.
This ratio is consistent with the (lower range of the) model predictions 
of Treister et al. (2004). A
significant number of additional obscured AGN may eventually be revealed
by {\it Spitzer} observations of X-ray quiet radio galaxies (Donley et al. 2005)
and by further characterization of ULIRGs harboring AGN but with cool IR
SEDs probably dominated by emission due to young stars.

\smallskip

We are grateful to the referee for comments that helped improve the paper.
We thank Belinda Wilkes and Joanna Kuraszkiewicz for interesting discussions. 
AAH acknowledges support from 
the Spanish Programa Nacional de Astronom\'{\i}a y Astrof\'{\i}sica 
under grant AYA2002-01055.
This work is based [in part] 
on observations made with the Spitzer Space Telescope, which 
is operated by the Jet Propulsion Laboratory, Caltech 
under NASA contract 1407. Support 
for this work was provided by NASA through Contract no. 
960785 and 1256790 issued by JPL/Caltech.
This research has made use of the NASA/IPAC Extragalactic Database (NED) 
which is operated by the Jet Propulsion Laboratory, California 
Institute of Technology, under contract with the National 
Aeronautics and Space Administration.

\appendix

\section*{Selection of the Sample of IR Power-Law Galaxies}


\begin{figure}
\epsscale{0.8}
\plotone{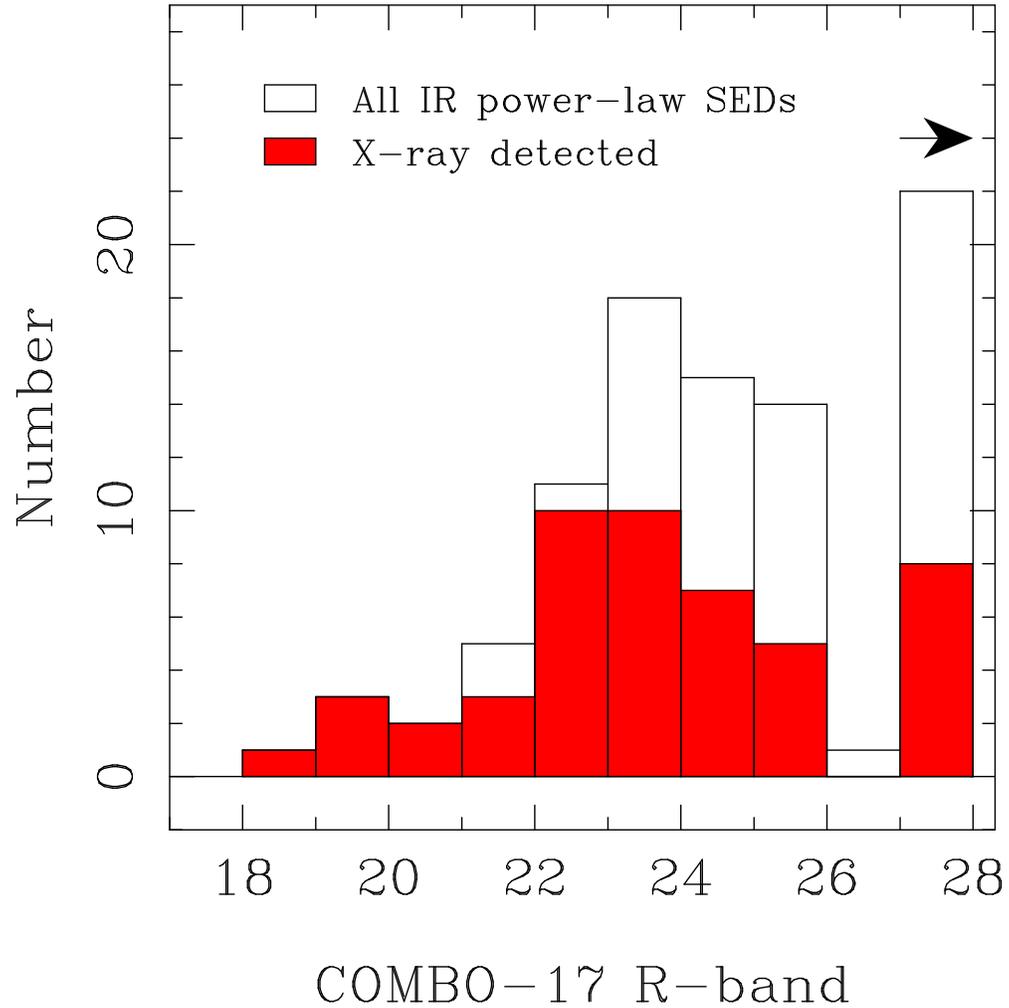}
\caption{Distributions of the COMBO-17 $R$-band
magnitudes (Wolf et al. 2004) 
of the 92 IR power-law SED galaxies in our
sample (empty histogram). The filled histogram
is galaxies detected in at least 
one of the {\it Chandra} X-ray bands. Those galaxies in our sample
not detected by COMBO-17 are displayed at an upper limit of $R=27\,$mag.}
\vspace{0.4cm}
\end{figure}

\subsection*{A.1 Optical and Near-Infrared Imaging and Band Merging}

The {\it Spitzer} images have been supplemented with 
optical and near-IR datasets
available from the literature: 1.) optical images ($UU_pBVRI$) from the ESO Imaging Survey 
(EIS, Arnouts et al. 2002); 2.) optical fluxes from the COMBO-17 survey 
(Wolf et al. 2004); 3.) $RIz$  imaging obtained by
Las Campanas Infrared Survey (Marzke et al. 1999); 4.) the 
{\it HST}/ACS $BViz$ and ground-based $JHK$ observations
from the GOODS team (Giavalisco 
et al. 2004); 5.) the near-IR $JK$ data released by
the EIS Deep Infrared Survey (Vandame et al. 2001); 
and 6.) the $I$-band photometry and
spectroscopic redshifts published by the VVDS  
(Le F\`evre et al. 2004).

For the imaging data, the source detection and photometry were carried out
with {\sc sextractor}. 
Merged catalogs with a $2\arcsec$
search radius in all the available bands were built by matching the
coordinates of the $24\,\mu$m sources to the deepest reference
optical band, in our case the $B$-band.  The fraction of multiple counterparts
within the $2\arcsec$ search radius was less than $3-5\%$ for the ground-based
data and less than $15\%$ for the {\it HST} data.
The depth of the observations and 
the procedure for obtaining photometry are
described in full detail by P\'erez-Gonz\'alez et al. (2005). The procedure
allowed us to obtain integrated fluxes for each filter in matched
apertures, and properly estimate colors for each source. For
sources not detected in the reference image, we used other optical/near-IR
images as the reference (if possible). In the case of the IRAC and
MIPS bands, the integrated flux was assumed to be that obtained from
PSF fitting.

\subsection*{A.2 Final Sample}

To obtain the final sample, the complete optical-to-24$\mu$m 
SEDs were visually inspected to 
eliminate any possible outliers. The criteria are not
uniform because of varying signal to noise from one object to another. The
number of outliers was less than $15\%$, and were mainly galaxies with bad
SEDs or galaxies resembling normal galaxies (see A.4 and Fig.~12). 
The final sample is composed of 92 galaxies and will be referred to as IR power-law
galaxies. They have been selected without regard to X-ray properties and therefore
should be unbiased in that regard. 
Table~1 gives  {\it Spitzer} coordinates, flux densities and
uncertainties, as well as the fitted IRAC spectral indices $\alpha$ for the
sample of IR power-law galaxies.

Fig.~10 shows the distribution of COMBO-17 $R$-band magnitudes (Wolf
et al. 2004) for our sample of galaxies compared to those 
not detected in X-rays. Galaxies not detected 
by COMBO-17 (22 out of 92) are shown as upper limits at $R=27$. It is clear 
from this figure that the fraction of IR
power-law galaxies not detected 
in X-rays increases towards fainter $R$-band magnitudes.
For comparison the candidate AGN not
present in the SDSS selected by Lacy et al. (2004) using 
IRAC colors are optically brighter than our sample, with 
$R$-band magnitudes in the range $R=18.5-22.2\,$mag. 

We note that only AGN dominated 
galaxies will be identified with this method, and that the large fraction of 
X-ray identified AGN in cosmological surveys with optical to mid-IR SEDs  
dominated by stellar emission (Alonso-Herrero et al. 2004) will
not be included in our sample.

\subsection*{A.3 Comparison with local Universe LIRGs and ULIRGs}
Since our sample of IR power-law galaxies is selected at $24\,\mu$m, and thus
is very likely to contain a large fraction of high redshift 
LIRGs and ULIRGs, it is of interest to simulate the expected SEDs using
local Universe analogs. 

Warm ULIRGs 
account for between 15 and 30\% of galaxies with 
$L_{\rm IR} > 10^{12}\,{\rm L}_\odot$
in the Bright Galaxy Sample (Sanders 
et al. 1988). The fraction of IR luminous
galaxies (LIRGs,
$10^{11}\,{\rm L}_\odot < L_{\rm IR} < 10^{12}\,{\rm L}_\odot$) 
and ULIRGs (cool and warm) containing an AGN increases with IR 
luminosity  (Veilleux et al. 1995, 1999, Tran et al. 2001); at IR luminosities
above L$_{\rm IR}> 10^{12.3}{\rm L}_\odot$ this fraction is over 
50\%. Intense star formation in dusty galaxies, on the other hand,  can also 
produce a bright mid-IR continuum that increases with wavelength, although
the stellar bump at $1.6\,\mu$m should still be present and prominent.
As shown by Le Floc'h et al. (2004) and 
Egami et al. (2004), when present in high-$z$ luminous
IR galaxies, the $1.6\,\mu$m rest-frame stellar 
bump is easily recognized using optical to mid-IR SEDs obtained virtually
identically to those used here (that is, a combination of IRAC four-color
photometry with data in the optical and near-IR). 
\begin{figure*}
\epsscale{0.3}
\plotone{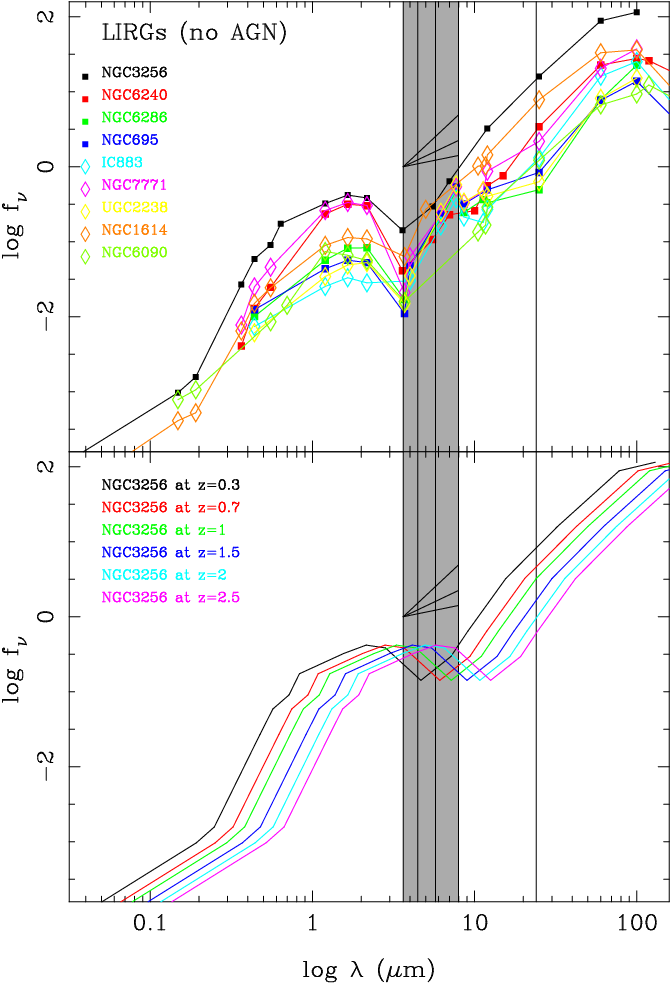}
\plotone{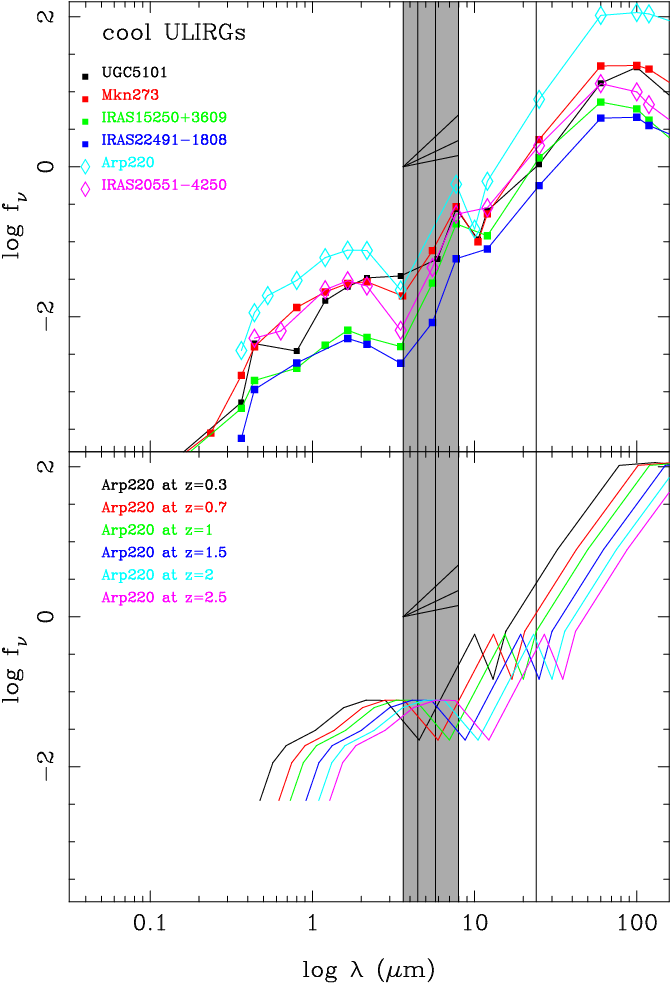}
\plotone{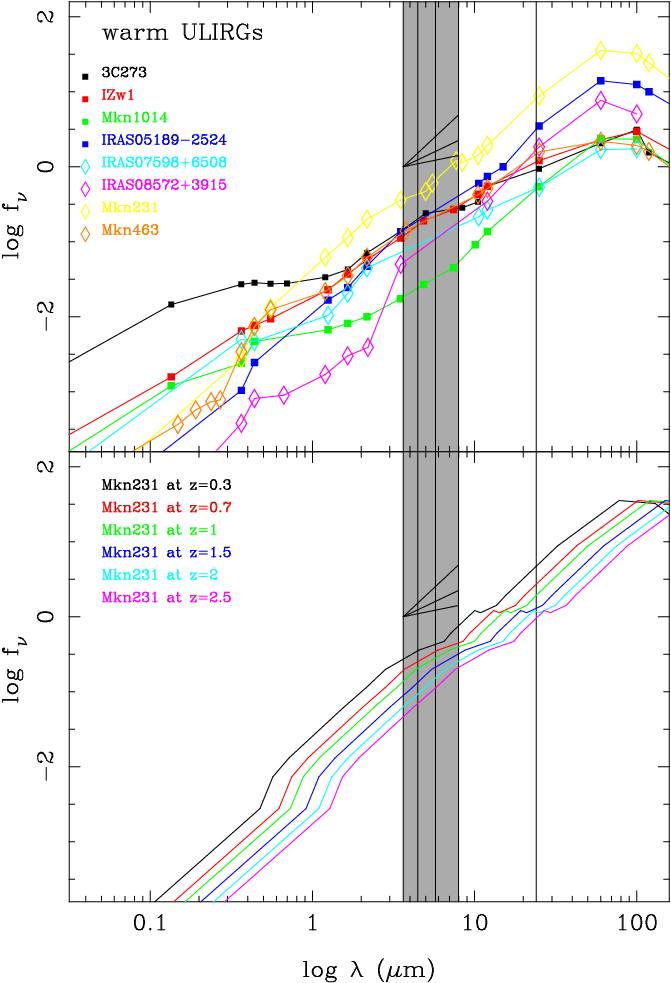}
\caption{The upper panels from left to right are SEDs of local Universe LIRGs 
with no evidence for the presence of an AGN, cool ULIRGs, 
and warm ULIRGs, respectively. The lower panels show one SED of each type 
as it would be observed for galaxies at different redshifts. We show 
the spectral range covered by the IRAC bands (shaded area), and the MIPS $24\,\mu$m band. 
We have also plotted power laws ($f_\nu \propto \nu^\alpha$) in the IRAC spectral range with 
spectral indices $\alpha = -0.5, \, -0.7\, -1.0$.}
\vspace{0.4cm}
\end{figure*}

To determine whether star-formation-dominated LIRGs and ULIRGs 
could be included in our sample, we have compiled data for 
local Universe LIRGs with no evidence for the presence of an AGN, 
as well as cool ULIRGs. Some of the galaxies in the latter category are
known to host a Seyfert nucleus in their centers (e.g., Mrk~273, 
UGC~5101). For comparison we have
also compiled data for  the warm ULIRGs in Sanders et al. 
(1988) detected in all four IRAS bands. These warm ULIRGs are
all known to contain a luminous AGN. The vast majority of the data have been taken
from the NASA/IPAC Extragalactic Database
(NED). We required fairly complete SEDs to include 
galaxies in this comparison, in particular in the 
near and mid-IR ranges. In addition to the data from NED, we
have made use of {\it ISO} data for non-AGN LIRGs from Lu et al. (2003) and 
for cool ULIRGs from Farrah et al. (2003), as well as optical and 
near-IR data from Surace, Sanders, \& Evans (2000). The results are
shown in Fig.~11. 

As expected (upper left panel of
Fig.~11), local LIRGs with no AGN show a very pronounced stellar bump
at $1.6\,\mu$m whereas this bump is absent or almost absent in warm ULIRGs
(upper right panel of Fig.~11). Cool
ULIRGs (middle panel of Fig.~11), 
even those containing an AGN (i.e., UGC~5101 and Mrk~273), represent 
intermediate cases. Also for those galaxies with available 
{\it ISO}/PHOT-S $7.7\,\mu$m spectroscopic data (Lutz et al. 1998;
Rigopoulou et al. 1999), 
we can see that polycyclic aromatic 
hydrocarbon (PAH) emission tends to be present in those galaxies with a strong
stellar continuum. Warm ULIRGs (see also SEDs in Sanders et al. 1988) 
show a range of spectral shapes, going from the typical shape of an optically 
selected QSO (IZw1) or radio loud QSO (3C273), to AGN with a more obscured 
UV and optical 
continuum (e.g., Mrk~463 and Mrk~231). 

In the lower panels of Fig.~11, 
for a representative object of each of these classes 
we have redshifted the SEDs to the range of redshifts we may expect for 
galaxies in our sample. We also show for comparison purposes 
power laws in the IRAC spectral range with 
spectral indices $\alpha = -0.5, \, -0.7\, -1.0$.
\smallskip

\begin{figure}
\epsscale{0.5}
\plotone{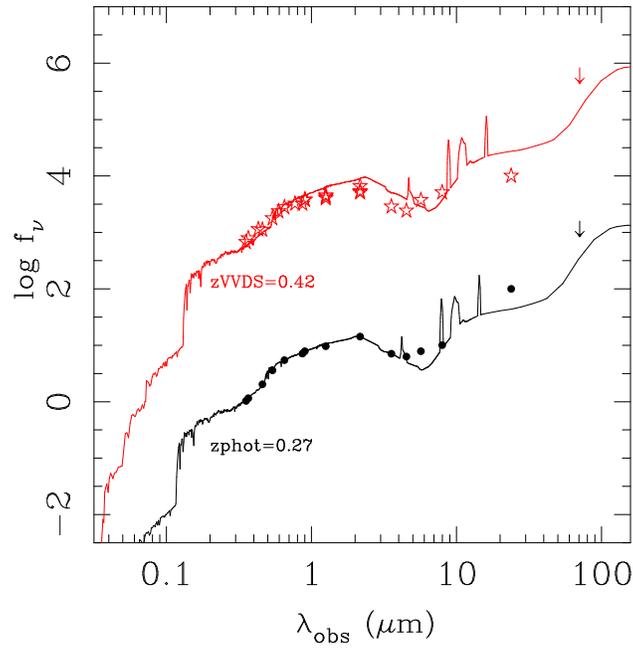}
\caption{Two nearby galaxies initially selected with the IR power-law 
criteria that show SEDs (filled dots and stars, arbitrarily shifted for
clarity) consistent with that of a normal (non-active) 
galaxy. The lines are a theoretical template for an Sc galaxy from 
Devriendt, Guiderdoni, \& Sadat (1999). }
\vspace{0.4cm}
\end{figure} 

Fig.~11 illustrates a number of important points. 
First, nearby ($z\le 0.4-0.5$) IR bright galaxies could be 
included in our initial sample since their rest-frame $4-8\,\mu$m continuum rises steeply resembling
the shape of a power law. However, once the complete  
optical to mid-IR SEDs are constructed, 
this kind of galaxy is easily identified and therefore excluded from our final
sample (see next section).
Second, cool ULIRGs at $0.4<z<2$ containing a Seyfert nucleus but whose
near-IR continuum is  dominated by stellar 
emission (from either a quiescent population or a starburst), such as Mrk~273, 
will not be included in our final sample because the $1.6\,\mu$m stellar bump
will get redshifted into the IRAC bands.  
AGN whose IR SEDs are dominated by 
stellar emission might make up a significant
fraction of the AGN  identified in deep X-ray surveys 
(Alonso-Herrero et al. 2004; Franceschini et al. 
2005), and are not selected with our criteria.
Lastly, for galaxies at $z>2$ not showing a very steep ($\alpha > -1$) 
IRAC continuum we may not be able to distinguish between the 
SED of a warm and a cool ULIRG. In such cases, we will have to use 
other criteria (i.e., X-ray and IR luminosities) 
to determine whether they are likely to contain an AGN.

\subsection*{A.4 Normal Galaxies}

As an illustration, in Fig.~12 we show two of the galaxies eliminated from the
sample whose SEDs resemble that of a normal galaxy.
Both are two nearby ($z<0.4-0.5$) 
galaxies dominated by stellar emission but showing a
steeply rising rest-frame $4-8\,\mu$m continuum. Although they
passed screening for the initial sample, they are easily recognized
as {\it normal galaxies from the 
optical to mid-IR SED, as they show the typical SED of a spiral
galaxy with a prominent $1.6\,\mu$m stellar bump. }

Normal galaxies in our sample are also obvious because
they are optically bright with respect to their X-ray fluxes (or to
the upper limits to the fluxes --- see further
discussion in \S7.1).
Of the two galaxies shown in Fig.~12, one has a spectroscopic redshift 
from the VVDS, and for the second one we use our own  phot-z (see Fig.~12). 
We find that both galaxies have X-ray
upper limits to the {\it Chandra} full-band luminosity of 
$L_{\rm 0.5-8keV} \leq 10^{41}
\,{\rm erg\,s}^{-1}$, well 
below the typical luminosity of  a Seyfert nucleus.

{\rotate

\begin{table*}
\caption{{\it Spitzer} coordinates and flux densities, and fitted IRAC spectral
  indices 
of the IR power-law galaxy sample.}
\tiny

\begin{tabular}{lcccccccccccccccc}
\hline
Name & RA & Dec & 
\multicolumn{2}{c}{${24\,\mu{\rm m}}$}& 
\multicolumn{2}{c}{${3.6\,\mu{\rm m}}$} & 
\multicolumn{2}{c}{${4.5\,\mu{\rm m}}$} & 
\multicolumn{2}{c}{${5.8\,\mu{\rm m}}$} & 
\multicolumn{2}{c}{${8.0\,\mu{\rm m}}$} & 
\multicolumn{2}{c}{${70\,\mu{\rm m}}$}  &
$\alpha$ & $\Delta \alpha$ \\
& & & 
$f_\nu$ & $\Delta f_\nu$ & 
$f_\nu$ & $\Delta f_\nu$ & 
$f_\nu$ & $\Delta f_\nu$ & 
$f_\nu$ & $\Delta f_\nu$ & 
$f_\nu$ & $\Delta f_\nu$ & 
$f_\nu$ & $\Delta f_\nu$ \\
& & & 
$\mu$Jy & $\mu$Jy & 
$\mu$Jy & $\mu$Jy & 
$\mu$Jy & $\mu$Jy & 
$\mu$Jy & $\mu$Jy & 
$\mu$Jy & $\mu$Jy & 
$\mu$Jy & $\mu$Jy \\
(1) & (2) & (3) & (4)  & (5) & (6) & (7) & (8) &
(9) & (10) & (11) & (12) & (13) & (14) & (15) &
(16) & (17)\\

\hline

mips000243 &  53.2050 &  -27.9180 & 141.0 & 29.9 & 11.2 &  1.1 & 10.1 &  1.1 & 11.5 &  2.2 & 17.0 &  2.5 & --    &--    & -0.5 &  0.2\\
mips000309 &  53.1125 &  -27.6848 & 908.0 & 69.9 &352.0 & 32.5 &320.0 & 31.0 &408.0 & 43.3 &549.0 & 50.4 & --    &--    & -0.6 &  0.2\\
mips000348 &  53.2562 &  -27.6951 & 167.0 & 27.2 & 34.5 &  3.2 & 30.8 &  3.0 & 35.5 &  4.2 & 51.6 &  5.1 & --    &--    & -0.6 &  0.2\\
mips002310 &  53.0577 &  -27.9335 & 131.0 & 25.7 &  5.9 &  0.6 &  7.8 &  0.8 & 14.7 &  2.4 & 27.1 &  3.2 & --    &--    & -2.0 &  0.2\\
mips002707 &  53.2816 &  -27.8573 &  80.3 & 24.2 & 15.3 &  1.5 & 16.1 &  1.6 & 16.0 &  2.5 & 24.9 &  3.0 & --    &--    & -0.6 &  0.2\\
mips002775 &  53.2472 &  -27.8164 & 781.0 & 52.9 & 43.4 &  4.1 & 52.6 &  5.1 & 68.2 &  7.5 &137.0 & 12.7 & 9036.5 &2441.0 & -1.4 &  0.2\\
mips002886 &  53.1422 &  -27.9445 & 108.0 & 27.0 &  8.5 &  0.8 & 11.1 &  1.2 & 14.6 &  2.4 & 21.9 &  2.8 & --    &--    & -1.2 &  0.2\\
mips003108 &  53.1744 &  -27.8673 & 108.0 & 26.2 & 14.5 &  1.4 & 11.6 &  1.2 & 13.5 &  2.3 & 20.7 &  2.7 & 7345.1 &2168.0 & -0.5 &  0.2\\
mips003133 &  53.1953 &  -27.8559 & 145.0 & 25.9 & 12.6 &  1.2 & 13.4 &  1.4 & 18.3 &  2.7 & 19.9 &  2.7 & 6039.5 &2476.2 & -0.6 &  0.2\\
mips003149 &  53.0606 &  -27.8826 & 223.0 & 28.2 & 11.6 &  1.1 & 12.9 &  1.3 & 18.4 &  2.7 & 23.2 &  2.9 & 3427.7 &1968.8 & -0.9 &  0.2\\
mips003175 &  53.0316 &  -27.8705 & 264.0 & 35.2 & 18.0 &  1.7 & 20.0 &  2.0 & 27.2 &  3.4 & 40.2 &  4.2 & --    &--    & -1.0 &  0.2\\
mips003412 &  52.9687 &  -27.8385 &  87.1 & 27.6 & 14.8 &  1.4 & 13.8 &  1.4 & 14.9 &  2.4 & 28.2 &  3.2 & 3228.5 &1692.1 & -0.8 &  0.2\\
mips003445 &  52.9773 &  -27.8516 & 131.0 & 28.9 & 13.1 &  1.3 & 13.8 &  1.4 & 19.5 &  2.8 & 25.6 &  3.0 & --    &--    & -0.9 &  0.2\\
mips003467 &  52.9601 &  -27.8701 & 499.0 & 34.0 & 27.1 &  2.5 & 33.3 &  3.3 & 55.8 &  6.2 & 99.6 &  9.4 & --    &--    & -1.7 &  0.2\\
mips003485 &  53.1053 &  -27.8750 & 136.0 & 23.2 &  9.3 &  0.9 & 12.5 &  1.3 & 20.3 &  2.9 & 26.6 &  3.1 & 3176.9 &1976.9 & -1.4 &  0.2\\
mips003528 &  53.1436 &  -27.8348 & 178.0 & 29.3 &  8.5 &  0.8 &  9.3 &  1.0 & 14.7 &  2.4 & 16.8 &  2.5 & 4655.9 &1940.8 & -0.9 &  0.2\\
mips003554 &  53.0161 &  -27.8915 &  76.4 & 26.3 & 14.4 &  1.4 & 14.7 &  1.5 & 15.0 &  2.5 & 23.2 &  2.9 & --    &--    & -0.6 &  0.2\\
mips003559 &  53.0025 &  -27.8982 & 149.0 & 34.4 &  8.2 &  0.8 & 10.1 &  1.1 & 18.4 &  2.7 & 34.4 &  3.7 & --    &--    & -1.9 &  0.2\\
mips003618 &  53.1489 &  -27.8212 & 519.0 & 41.0 & 14.9 &  1.4 & 16.4 &  1.6 & 30.3 &  3.7 & 36.8 &  3.9 & 3318.9 &1827.6 & -1.2 &  0.2\\
mips003636 &  52.9666 &  -27.8909 & 298.0 & 50.7 &  6.4 &  0.7 &  9.6 &  1.0 & 18.1 &  2.7 & 34.7 &  3.7 & --    &--    & -2.1 &  0.2\\
mips003674 &  52.9928 &  -27.8451 & 833.0 & 50.7 & 23.7 &  2.2 & 25.1 &  2.5 & 42.2 &  4.9 & 86.5 &  8.2 & 8871.6 &1782.9 & -1.7 &  0.2\\
mips003735 &  53.2866 &  -27.7151 & 164.0 & 33.0 & 26.1 &  2.4 & 25.9 &  2.6 & 31.1 &  3.8 & 41.9 &  4.3 & --    &--    & -0.6 &  0.2\\
mips003777 &  52.9372 &  -27.8605 & 219.0 & 33.4 & 13.1 &  1.3 & 14.3 &  1.4 & 18.7 &  2.7 & 17.9 &  2.5 & --    &--    & -0.5 &  0.2\\
mips003888 &  53.1151 &  -27.6959 & 351.0 & 36.4 & 46.0 &  4.3 & 39.7 &  3.9 & 55.7 &  6.2 & 74.9 &  7.2 & --    &--    & -0.7 &  0.2\\
mips003915 &  53.1589 &  -27.6625 & 559.0 & 50.6 & 76.0 &  7.1 & 72.5 &  7.1 & 91.6 &  9.9 &137.0 & 12.7 & 4217.0 &2434.7 & -0.8 &  0.2\\
mips003938 &  53.1310 &  -27.7732 & 328.0 & 37.2 & 23.8 &  2.2 & 24.5 &  2.4 & 42.6 &  4.9 & 59.7 &  5.8 & 6792.0 &2147.3 & -1.2 &  0.2\\
mips004254 &  52.9558 &  -27.7763 &  91.1 & 22.6 & 14.8 &  1.4 & 15.2 &  1.5 & 20.1 &  2.8 & 35.8 &  3.8 & 4786.3 &2350.8 & -1.1 &  0.2\\
mips004258 &  53.0455 &  -27.7375 & 286.0 & 33.9 & 35.6 &  3.3 & 45.9 &  4.5 & 56.6 &  6.3 & 75.0 &  7.2 & 3104.6 &2163.5 & -0.9 &  0.2\\
mips004281 &  53.0334 &  -27.7110 & 511.0 & 45.1 & 23.9 &  2.2 & 21.7 &  2.2 & 34.5 &  4.1 & 54.8 &  5.4 & --    &--    & -1.1 &  0.2\\
mips005415 &  53.0167 &  -27.6238 &  85.5 & 27.0 & 17.5 &  1.7 & 16.0 &  1.6 & 22.6 &  3.0 & 28.0 &  3.2 & --    &--    & -0.7 &  0.2\\
mips009708 &  53.1802 &  -27.8206 & 144.0 & 24.6 & 18.1 &  1.7 & 19.1 &  1.9 & 29.0 &  3.6 & 28.0 &  3.2 & --    &--    & -0.6 &  0.2\\
mips009714 &  52.9470 &  -27.8871 & 214.0 & 30.6 & 17.6 &  1.7 & 23.5 &  2.3 & 39.7 &  4.6 & 67.4 &  6.5 & --    &--    & -1.7 &  0.2\\
mips009757 &  53.1929 &  -27.7755 &  89.9 & 23.5 &  6.8 &  0.7 &  6.3 &  0.7 & 11.0 &  2.2 & 14.6 &  2.3 & 3872.6 &2061.0 & -1.0 &  0.2\\
mips010697 &  53.1573 &  -27.8701 &1010.0 & 62.1 & 47.1 &  4.4 & 66.3 &  6.5 &101.0 & 10.9 &217.0 & 20.0 & --    &--    & -1.9 &  0.2\\
mips010760 &  53.0821 &  -27.7673 & 275.0 & 37.1 & 10.9 &  1.1 & 13.4 &  1.4 & 20.5 &  2.9 & 18.2 &  2.5 & 4497.8 &1689.6 & -0.8 &  0.2\\
mips010785 &  53.1071 &  -27.7183 & 480.0 & 45.3 & 14.1 &  1.4 & 21.8 &  2.2 & 53.8 &  6.0 &125.0 & 11.6 & --    &--    & -2.8 &  0.2\\
mips010787 &  53.1049 &  -27.7053 & 499.0 & 40.0 & 52.0 &  4.8 & 61.4 &  6.0 & 69.0 &  7.6 & 94.7 &  8.9 & --    &--    & -0.7 &  0.2\\
mips010823 &  53.1110 &  -27.6702 & 116.0 & 29.5 & 13.4 &  1.3 & 14.7 &  1.5 & 15.7 &  2.5 & 32.7 &  3.6 & --    &--    & -1.1 &  0.2\\
mips012254 &  53.2494 &  -27.7967 & 101.0 & 29.9 & 16.9 &  1.6 & 17.3 &  1.7 & 23.7 &  3.1 & 39.7 &  4.1 & --    &--    & -1.1 &  0.2\\
mips012297 &  53.1995 &  -27.7092 & 995.0 & 51.3 &140.0 & 13.0 &130.0 & 12.6 &155.0 & 16.6 &220.0 & 20.3 & --    &--    & -0.6 &  0.2\\
mips012310 &  53.0222 &  -27.7792 & 182.0 & 55.2 & 12.0 &  1.2 & 12.0 &  1.2 & 12.3 &  2.3 & 23.0 &  2.9 & 8669.6 &2219.6 & -0.8 &  0.2\\
mips012314 &  52.9680 &  -27.7981 & 369.0 & 65.2 & 18.3 &  1.7 & 19.9 &  2.0 & 33.0 &  4.0 & 40.9 &  4.2 & 6823.4 &1936.8 & -1.1 &  0.2\\
mips012329 &  53.2050 &  -27.6808 &2100.0 & 83.4 &112.0 & 10.3 &173.0 & 16.8 &293.0 & 31.1 &489.0 & 44.9 &10447.2 &2050.5 & -1.9 &  0.2\\
mips012371 &  53.0676 &  -27.6585 &1350.0 & 74.5 &140.0 & 13.0 &170.0 & 16.5 &241.0 & 25.6 &390.0 & 35.8 & --    &--    & -1.3 &  0.2\\
mips012940 &  53.2571 &  -27.9719 & 655.0 & 50.5 & 67.4 &  6.3 & 72.5 &  7.1 &109.0 & 11.7 &188.0 & 17.4 & 3184.2 &2188.5 & -1.3 &  0.2\\
mips013087 &  53.1104 &  -27.6766 &2060.0 & 94.5 &168.0 & 15.5 &176.0 & 17.1 &256.0 & 27.2 &449.0 & 41.3 & 5296.6 &2170.8 & -1.3 &  0.2\\
mips013611 &  53.0361 &  -27.7930 &3320.0 &110.0 &410.0 & 37.8 &406.0 & 39.5 &564.0 & 59.7 &734.0 & 67.2 & --    &--    & -0.8 &  0.2\\
mips014534 &  53.0061 &  -27.6940 & 234.0 & 51.0 & 24.7 &  2.3 & 26.6 &  2.6 & 31.7 &  3.9 & 50.1 &  5.0 & --    &--    & -0.9 &  0.2\\
mips014635 &  53.1250 &  -27.7585 & 442.0 & 59.2 & 66.4 &  6.2 & 64.5 &  6.3 & 69.9 &  7.7 &118.0 & 11.1 & --    &--    & -0.7 &  0.2\\

\hline
\end{tabular}
\end{table*}

\newpage
\begin{table*}
\setcounter{table}{0}
\caption{Continued.}
\tiny
\begin{tabular}{lcccccccccccccccc}
\hline
Name & RA & Dec & 
\multicolumn{2}{c}{${24\,\mu{\rm m}}$}& 
\multicolumn{2}{c}{${3.6\,\mu{\rm m}}$} & 
\multicolumn{2}{c}{${4.5\,\mu{\rm m}}$} & 
\multicolumn{2}{c}{${5.8\,\mu{\rm m}}$} & 
\multicolumn{2}{c}{${8.0\,\mu{\rm m}}$} & 
\multicolumn{2}{c}{${70\,\mu{\rm m}}$}  &
$\alpha$ & $\Delta \alpha$ \\
\\
& & & 
$f_\nu$ & $\Delta f_\nu$ & 
$f_\nu$ & $\Delta f_\nu$ & 
$f_\nu$ & $\Delta f_\nu$ & 
$f_\nu$ & $\Delta f_\nu$ & 
$f_\nu$ & $\Delta f_\nu$ & 
$f_\nu$ & $\Delta f_\nu$ \\
& & & 
$\mu$Jy & $\mu$Jy & 
$\mu$Jy & $\mu$Jy & 
$\mu$Jy & $\mu$Jy & 
$\mu$Jy & $\mu$Jy & 
$\mu$Jy & $\mu$Jy & 
$\mu$Jy & $\mu$Jy \\
(1) & (2) & (3) & (4)  & (5) & (6) & (7) & (8) &
(9) & (10) & (11) & (12) & (13) & (14) & (15) &
(16) & (17)\\

\hline

mips000314 &  53.0968 &  -27.6725 & 109.0 & 26.5 &  6.3 &  0.7 &  8.2 &  0.9 &  9.9 &  2.1 & 12.3 &  2.2 & 2500.3 &1775.8 & -0.9 &  0.3\\
mips002219 &  53.1311 &  -27.9810 & 252.0 & 31.6 & 10.4 &  1.0 & 12.7 &  1.3 & 12.9 &  2.3 & 18.8 &  2.6 & 3243.4 &2333.4 & -0.7 &  0.2\\
mips002556 &  53.1769 &  -27.9759 &  75.4 & 22.3 &  3.9 &  0.4 &  5.5 &  0.6 &  9.6 &  2.1 & 20.0 &  2.7 & 2985.4 &2140.1 & -2.1 &  0.2\\
mips002581 &  53.1458 &  -27.9871 & 174.0 & 33.6 & 11.2 &  1.1 & 11.2 &  1.2 & 11.1 &  2.2 & 22.3 &  2.8 & 3655.9 &2526.6 & -0.8 &  0.2\\
mips002613 &  53.2175 &  -27.9398 & 107.0 & 26.2 & 11.6 &  1.1 & 12.0 &  1.2 & 18.6 &  2.7 & 16.0 &  2.4 & --    &--    & -0.5 &  0.2\\
mips002788 &  53.3056 &  -27.8057 & 172.0 & 30.3 &  6.7 &  0.7 &  7.6 &  0.8 & 10.2 &  2.1 & 10.2 &  2.1 & 3944.6 &2395.0 & -0.6 &  0.3\\
mips002957 &  53.0709 &  -27.9432 &  69.2 & 23.6 &  5.8 &  0.6 &  9.1 &  1.0 & 11.9 &  2.2 & 16.7 &  2.5 & 5847.9 &1799.9 & -1.3 &  0.2\\
mips003068 &  53.2262 &  -27.8589 &  53.0 & 24.5 &  3.0 &  0.4 &  3.1 &  0.4 &  8.1 &  2.0 & 12.7 &  2.2 & --    &--    & -1.9 &  0.3\\
mips003181 &  53.0701 &  -27.8611 & 132.0 & 30.5 &  8.4 &  0.8 & 10.3 &  1.1 & 10.4 &  2.2 & 12.4 &  2.2 & --    &--    & -0.5 &  0.3\\
mips003182 &  53.0384 &  -27.8622 & 352.0 & 31.2 & 34.4 &  3.2 & 34.8 &  3.4 & 41.9 &  4.8 & 55.8 &  5.5 & --    &--    & -0.6 &  0.2\\
mips003271 &  52.8992 &  -27.8597 &  79.7 & 36.1 & 18.5 &  1.8 & 19.8 &  2.0 & 29.5 &  3.7 & 36.6 &  3.9 & --    &--    & -0.9 &  0.2\\
mips003357 &  52.9408 &  -27.8085 & 103.0 & 24.1 &  5.5 &  0.6 &  6.0 &  0.7 & 11.8 &  2.2 & 14.6 &  2.3 & 2760.6 &1738.4 & -1.3 &  0.2\\
mips003363 &  52.9462 &  -27.8194 & 103.0 & 29.2 & 15.2 &  1.4 & 14.4 &  1.5 & 12.2 &  2.3 & 23.2 &  2.9 & --    &--    & -0.5 &  0.2\\
mips003377 &  52.9609 &  -27.8162 & 133.0 & 29.9 &  9.8 &  1.0 & 10.4 &  1.1 & 17.7 &  2.6 & 16.7 &  2.4 & --    &--    & -0.8 &  0.2\\
mips003395 &  52.9728 &  -27.8185 &  81.6 & 28.3 &  3.5 &  0.4 &  4.9 &  0.6 & 10.1 &  2.1 & 11.7 &  2.2 & --    &--    & -1.6 &  0.3\\
mips003454 &  52.9504 &  -27.8540 & 210.0 & 28.8 & 12.6 &  1.2 & 13.2 &  1.3 & 14.9 &  2.4 & 18.8 &  2.6 & --    &--    & -0.5 &  0.2\\
mips003480 &  53.0113 &  -27.9150 & 127.0 & 33.3 &  4.6 &  0.5 &  5.3 &  0.6 &  8.2 &  2.0 & 11.7 &  2.2 & 7112.1 &2059.1 & -1.2 &  0.3\\
mips003537 &  53.1698 &  -27.8241 & 214.0 & 31.2 & 10.4 &  1.0 & 11.7 &  1.2 & 18.0 &  2.7 & 15.0 &  2.4 & --    &--    & -0.6 &  0.2\\
mips003556 &  53.0141 &  -27.8873 & 262.0 & 34.2 &  7.8 &  0.8 &  7.1 &  0.8 & 12.9 &  2.3 & 17.3 &  2.5 & 5296.6 &2140.9 & -1.0 &  0.2\\
mips003639 &  53.0058 &  -27.8737 & 236.0 & 27.2 & 13.3 &  1.3 & 14.6 &  1.5 & 17.2 &  2.6 & 26.0 &  3.1 & --    &--    & -0.8 &  0.2\\
mips003641 &  53.0014 &  -27.8653 & 114.0 & 24.5 &  9.5 &  0.9 &  8.0 &  0.9 & 11.4 &  2.2 & 16.2 &  2.4 & --    &--    & -0.7 &  0.2\\
mips003832 &  53.1983 &  -27.7478 & 272.0 & 33.3 & 13.4 &  1.3 & 15.2 &  1.5 & 16.0 &  2.5 & 21.0 &  2.7 & --    &--    & -0.5 &  0.2\\
mips003871 &  53.1379 &  -27.7002 & 183.0 & 35.1 & 12.4 &  1.2 & 10.4 &  1.1 & 15.0 &  2.5 & 17.9 &  2.5 & 7145.0 &2201.8 & -0.5 &  0.2\\
mips004100 &  52.9933 &  -27.7934 &  95.9 & 24.5 &  4.7 &  0.5 &  4.3 &  0.5 &  5.9 &  1.9 & 10.7 &  2.1 & --    &--    & -0.9 &  0.3\\
mips004104 &  52.9897 &  -27.7513 & 146.0 & 31.2 & 13.0 &  1.2 & 13.6 &  1.4 & 23.6 &  3.1 & 21.2 &  2.7 & --    &--    & -0.8 &  0.2\\
mips004127 &  52.9616 &  -27.7844 & 369.0 & 40.8 & 17.0 &  1.6 & 17.5 &  1.7 & 17.2 &  2.6 & 30.4 &  3.4 & 3006.1 &2282.5 & -0.7 &  0.2\\
mips004145 &  53.2298 &  -27.7087 & 177.0 & 43.9 &  5.6 &  0.6 &  7.8 &  0.8 & 10.5 &  2.2 & 22.0 &  2.8 & --    &--    & -1.7 &  0.2\\
mips004158 &  53.2556 &  -27.6788 & 596.0 & 70.7 & 63.7 &  5.9 & 63.2 &  6.2 & 77.7 &  8.5 &121.0 & 11.2 & 6698.8 &2231.5 & -0.8 &  0.2\\
mips004196 &  53.0794 &  -27.7416 & 253.0 & 28.5 &  6.8 &  0.7 & 10.7 &  1.1 & 22.7 &  3.1 & 35.9 &  3.8 & --    &--    & -2.1 &  0.2\\
mips004244 &  52.9554 &  -27.7692 & 195.0 & 31.3 & 13.8 &  1.3 & 13.3 &  1.4 & 24.0 &  3.2 & 21.0 &  2.7 & --    &--    & -0.7 &  0.2\\
mips004982 &  52.9825 &  -27.6782 & 279.0 & 33.0 & 10.2 &  1.0 & 11.8 &  1.2 & 19.5 &  2.8 & 19.6 &  2.6 & 6652.7 &2234.0 & -0.9 &  0.2\\
mips005020 &  53.0775 &  -27.6349 & 289.0 & 37.2 &  5.4 &  0.6 &  7.1 &  0.8 & 11.4 &  2.2 & 22.7 &  2.8 & 4775.3 &2040.7 & -1.8 &  0.2\\
mips005025 &  52.9845 &  -27.6672 & 215.0 & 28.7 & 17.1 &  1.6 & 19.5 &  1.9 & 27.9 &  3.5 & 25.0 &  3.0 & --    &--    & -0.6 &  0.2\\
mips005381 &  53.0331 &  -27.6256 & 946.0 & 60.1 & 49.4 &  4.6 & 54.3 &  5.3 & 89.7 &  9.7 &173.0 & 16.1 & --    &--    & -1.6 &  0.2\\
mips009632 &  53.2606 &  -27.9459 & 255.0 & 31.8 & 11.4 &  1.1 & 11.7 &  1.2 & 13.1 &  2.3 & 19.1 &  2.6 & --    &--    & -0.6 &  0.2\\
mips009713 &  52.9383 &  -27.8845 & 180.0 & 31.0 &  8.1 &  0.8 &  9.2 &  1.0 & 11.2 &  2.2 & 15.1 &  2.4 & --    &--    & -0.8 &  0.2\\
mips009742 &  53.2202 &  -27.7456 & 119.0 & 28.7 &  6.5 &  0.7 &  6.4 &  0.7 & 10.6 &  2.2 & 10.4 &  2.1 & --    &--    & -0.7 &  0.3\\
mips009751 &  53.2235 &  -27.7151 & 209.0 & 30.2 &  8.0 &  0.8 &  8.3 &  0.9 & 15.1 &  2.5 & 13.7 &  2.3 & --    &--    & -0.8 &  0.2\\
mips009834 &  53.1158 &  -27.6306 &1120.0 & 73.6 &157.0 & 14.5 &175.0 & 17.0 &209.0 & 22.2 &306.0 & 28.2 & 6823.4 &2012.8 & -0.8 &  0.2\\
mips010694 &  53.1030 &  -27.8930 & 241.0 & 37.5 &  9.0 &  0.9 & 10.2 &  1.1 & 12.6 &  2.3 & 19.1 &  2.6 & --    &--    & -0.9 &  0.2\\
mips010818 &  53.1134 &  -27.6811 & 171.0 & 32.8 &  8.9 &  0.9 & 10.3 &  1.1 & 17.1 &  2.6 & 17.1 &  2.5 & 5675.4 &2204.1 & -0.9 &  0.2\\
mips013065 &  53.1754 &  -27.6949 & 146.0 & 26.7 &  9.2 &  0.9 & 12.8 &  1.3 & 15.0 &  2.5 & 21.9 &  2.8 & 8072.4 &2320.3 & -1.1 &  0.2\\
mips013589 &  52.9784 &  -27.8949 & 273.0 & 41.3 & 14.3 &  1.4 & 15.6 &  1.6 & 24.1 &  3.2 & 20.9 &  2.7 & --    &--    & -0.6 &  0.2\\

\hline

\end{tabular}
Notes.--- Column~(1): MIPS $24\,\mu$m name. Columns~(2) and (3): MIPS $24\,\mu$m
J2000.0 RA and Dec. Columns~(4) and (5): MIPS $24\,\mu$m flux density 
and error. 
Columns~(6) through (13): IRAC flux densities and errors. 
Columns~(14) and (15): MIPS $70\,\mu$m flux density and
error. Columns~(16) and (17): IRAC spectral index and uncertainty.
\end{table*}

}

\begin{table*}
\caption{X-ray properties of the IR power-law galaxy sample.}
\tiny
\begin{tabular}{lcccccccc}
\hline
Name & $f_{0.5-8{\rm keV}}$ & $f_{0.5-2{\rm keV}}$ &
$f_{2-8{\rm keV}}$ & $\Gamma$ & off-axis angle & ID\\
& erg cm$^{-2}$ s$^{-1}$  & erg cm$^{-2}$ s$^{-1}$ & erg cm$^{-2}$ s$^{-1}$\\ 
(1) & (2) & (3) & (4)  & (5) & (6) & (7)\\
\hline
mips000243 &0.2650E-14 &0.7800E-15 &0.1900E-14&   1.39&   7.95 &A277 \\
mips000309 &0.1210E-12 &0.4940E-13 &0.6860E-13&   1.79&   7.53 &A177 \\
mips000348 &0.1090E-13 &0.1530E-14 &0.9400E-14&   0.73&  10.09 &A310 \\
mips002310 &0.2400E-14 &0.3080E-15 &0.2050E-14&   0.67&   8.04 &A92  \\
mips002707 &0.1390E-13 &0.3970E-14 &0.9480E-14&   1.40&   9.16 &A322 \\
mips002775 &0.9670E-14 &0.3890E-15 &0.9890E-14&  -0.31&   6.89 &A302 \\
mips002886 &0.6430E-15 &-0.1390E-15 &-0.9690E-15&   1.40&   8.19 &S1   \\
mips003108 &0.5730E-14 &0.1670E-14 &0.3820E-14&   1.43&   4.57 &A251 \\
mips003133 &0.2270E-15 &-0.7880E-16 &-0.4530E-15&   1.40&   4.93 &S2   \\
mips003149 &0.1750E-15 &-0.6840E-16 &-0.4550E-15&   1.40&   5.29 &S3   \\
mips003175 &0.3320E-14 &0.5430E-15 &0.2760E-14&   0.86&   5.82 &A59  \\
mips003412 &0.1000E-13 &0.1610E-14 &0.8640E-14&   0.82&   8.07 &A16  \\
mips003445 &0.9000E-15 &0.1610E-15 &-0.8960E-15&   1.40&   7.83 &A22  \\
mips003467 &0.1580E-14 &-0.1340E-15 &0.1100E-14&   0.00&   9.10 &G319 \\
mips003485 &-0.1970E-15 &0.5320E-16 &-0.3220E-15&   1.40&   3.96 &A165 \\
mips003528 &-0.1730E-15 &0.4600E-16 &-0.3850E-15&   1.40&   2.01 &S4   \\
mips003554 &0.3640E-14 &0.1280E-14 &0.2170E-14&   1.65&   7.26 &A45  \\
mips003559 &0.4010E-14 &0.6380E-15 &0.3470E-14&   0.82&   8.07 &A36  \\
mips003618 &0.7760E-15 &0.5430E-16 &0.7010E-15&   0.19&   1.78 &A219 \\
mips003636 &0.1690E-14 &0.2830E-15 &-0.1330E-14&   0.92&   9.36 &A14  \\
mips003674 &0.1000E-13 &0.2240E-14 &0.7480E-14&   1.17&   6.94 &A30  \\
mips003735 &0.7910E-14 &0.3180E-14 &0.4750E-14&   1.74&  10.63 &A323 \\
mips003777 &0.1720E-14 &0.4210E-15 &-0.1420E-14&   1.16&  10.06 &A3   \\
mips003888 &0.1160E-13 &0.4880E-15 &0.1160E-13&  -0.25&   6.86 &A179 \\
mips003915 &0.3220E-13 &0.1600E-13 &0.1550E-13&   2.05&   9.12 &A234 \\
mips003938 &0.2480E-15 &0.6340E-16 &-0.3270E-15&   1.40&   2.34 &A197 \\
mips004254 &0.3950E-14 &0.1170E-14 &0.2730E-14&   1.42&   8.81 &A7   \\
mips004258 &0.1640E-13 &0.6140E-14 &0.9810E-14&   1.69&   5.80 &A76  \\
mips004281 &0.1450E-14 &0.2340E-15 &0.1190E-14&   0.86&   7.46 &A60  \\
mips005415 &0.2920E-14 &-0.1570E-14 &-0.6340E-14&   1.40&  12.39 &S7   \\
mips009708 &0.1010E-13 &0.3660E-14 &0.6110E-14&   1.66&   3.38 &A256 \\
mips009714 &0.1450E-13 &0.4000E-14 &0.1020E-13&   1.36&  10.15 &A4   \\
mips009757 &0.7030E-14 &0.1900E-15 &0.7260E-14&  -0.61&   4.52 &A268 \\
mips010697 &0.1520E-13 &0.5970E-14 &0.8780E-14&   1.75&   4.17 &A230 \\
mips010760 &-0.1450E-15 &0.3060E-16 &-0.2020E-15&   1.40&   3.22 &S8   \\
mips010785 &0.6120E-14 &0.1090E-14 &0.4940E-14&   0.94&   5.54 &A166 \\
mips010787 &0.6320E-14 &0.2790E-14 &0.3350E-14&   1.90&   6.33 &A163 \\
mips010823 &0.1010E-13 &0.2630E-14 &0.7030E-14&   1.32&   8.39 &A174 \\
mips012254 &0.1930E-13 &0.7420E-14 &0.1150E-13&   1.71&   7.04 &A305 \\
mips012297 &0.3790E-13 &0.7920E-14 &0.2960E-13&   1.08&   7.46 &A274 \\
mips012310 &0.3370E-14 &0.1200E-15 &0.3620E-14&  -0.43&   5.39 &A50  \\
mips012314 &0.2170E-14 &0.4930E-15 &0.1570E-14&   1.20&   7.97 &A15  \\
mips012329 &0.7060E-14 &-0.3650E-15 &0.8340E-14&  -0.23&   9.06 &A278 \\
mips012371 &0.3110E-13 &0.1260E-13 &0.1800E-13&   1.77&   9.48 &A109 \\
mips012940 &0.4610E-13 &0.1190E-13 &0.3110E-13&   1.34&  12.17 &A313 \\
mips013087 &0.2490E-13 &0.1280E-13 &0.1180E-13&   2.08&   8.02 &A173 \\
mips013611 &0.1150E-12 &0.4460E-13 &0.6770E-13&   1.73&   4.45 &A66  \\
mips014534 &0.6980E-14 &0.1620E-14 &0.5630E-14&   1.13&   9.14 &A38  \\
mips014635 &0.2130E-13 &0.7770E-14 &0.1290E-13&   1.66&   3.14 &A191 \\

\hline
\end{tabular}
\end{table*}

\begin{table*}
\setcounter{table}{1}
\caption{Continued.}
\tiny
\begin{tabular}{lcccccccc}
\hline
Name & $f_{0.5-8{\rm keV}}$ & $f_{0.5-2{\rm keV}}$ &
$f_{2-8{\rm keV}}$ & $\Gamma$ & off-axis angle & ID\\
& erg cm$^{-2}$ s$^{-1}$  & erg cm$^{-2}$ s$^{-1}$ & erg cm$^{-2}$ s$^{-1}$\\ 

(1) & (2) & (3) & (4)  & (5) & (6) & (7)\\

\hline

mips000314 &-0.4490E-15 &-0.1330E-15 &-0.7780E-15&   1.40&   8.33 &-    \\
mips002219 &-0.6670E-15 &-0.1970E-15 &-0.1150E-14&   1.40&  10.27 &-    \\
mips002556 &-0.9110E-15 &-0.2290E-15 &-0.1750E-14&   1.40&  10.43 &-    \\
mips002581 &-0.8460E-15 &-0.2110E-15 &-0.1530E-14&   1.40&  10.72 &-    \\
mips002613 &-0.6160E-15 &-0.1710E-15 &-0.1090E-14&   1.40&   9.41 &-    \\
mips002788 &-0.6640E-15 &-0.2010E-15 &-0.1150E-14&   1.40&   9.98 &-    \\
mips002957 &-0.4670E-15 &-0.1440E-15 &-0.8110E-15&   1.40&   8.36 &-    \\
mips003068 &-0.2590E-15 &-0.5460E-16 &-0.5220E-15&   1.40&   6.47 &-    \\
mips003181 &-0.1900E-15 &-0.5910E-16 &-0.3140E-15&   1.40&   3.96 &-    \\
mips003182 &-0.2330E-15 &-0.7700E-16 &-0.3620E-15&   1.40&   5.23 &-    \\
mips003271 &-0.7730E-14 &-0.9990E+03 &-0.1130E-13&   1.40&  11.96 &-    \\
mips003357 &-0.5570E-15 &-0.1350E-15 &-0.1040E-14&   1.40&   9.38 &-    \\
mips003363 &-0.5170E-15 &-0.1400E-15 &-0.9240E-15&   1.40&   9.11 &-    \\
mips003377 &-0.4600E-15 &-0.1340E-15 &-0.8290E-15&   1.40&   8.32 &-    \\
mips003395 &-0.3640E-15 &-0.9990E-16 &-0.6620E-15&   1.40&   7.70 &-    \\
mips003454 &-0.4800E-15 &-0.1170E-15 &-0.9310E-15&   1.40&   9.25 &-    \\
mips003480 &-0.4920E-15 &-0.1290E-15 &-0.9350E-15&   1.40&   8.45 &-    \\
mips003537 &-0.1290E-15 &-0.7300E-16 &-0.2660E-15&   1.40&   2.89 &-    \\
mips003556 &-0.3160E-15 &-0.8700E-16 &-0.5710E-15&   1.40&   7.18 &-    \\
mips003639 &-0.3170E-15 &-0.9600E-16 &-0.5440E-15&   1.40&   7.05 &-    \\
mips003641 &-0.3060E-15 &-0.8010E-16 &-0.5430E-15&   1.40&   7.00 &-    \\
mips003832 &-0.2160E-15 &-0.6230E-16 &-0.3990E-15&   1.40&   5.69 &-    \\
mips003871 &-0.3310E-15 &-0.9010E-16 &-0.5870E-15&   1.40&   6.69 &-    \\
mips004100 &-0.2690E-15 &-0.7040E-16 &-0.5090E-15&   1.40&   6.67 &-    \\
mips004104 &-0.3580E-15 &-0.9600E-16 &-0.6200E-15&   1.40&   7.65 &-    \\
mips004127 &-0.4720E-15 &-0.1240E-15 &-0.8560E-15&   1.40&   8.42 &-    \\
mips004145 &-0.7470E-15 &-0.2060E-15 &-0.1320E-14&   1.40&   8.52 &-    \\
mips004158 &-0.1510E-13 &-0.4140E-14 &-0.2670E-13&   1.40&  10.76 &-    \\
mips004196 &-0.2010E-15 &-0.6200E-16 &-0.3750E-15&   1.40&   4.58 &-    \\
mips004244 &-0.5780E-15 &-0.1750E-15 &-0.1010E-14&   1.40&   8.95 &-    \\
mips004982 &-0.1480E-14 &-0.4780E-15 &-0.2520E-14&   1.40&  10.68 &-    \\
mips005020 &-0.1830E-14 &-0.4990E-15 &-0.3090E-14&   1.40&  10.73 &-    \\
mips005025 &-0.1950E-14 &-0.4810E-15 &-0.3620E-14&   1.40&  11.11 &-    \\
mips005381 &-0.1430E-13 &-0.3330E-14 &-0.2560E-13&   1.40&  11.95 &-    \\
mips009632 &-0.2820E-14 &-0.7950E-15 &-0.4600E-14&   1.40&  11.14 &-    \\
mips009713 &-0.8010E-15 &-0.2180E-15 &-0.1480E-14&   1.40&  10.51 &-    \\
mips009742 &-0.3260E-15 &-0.7670E-16 &-0.6190E-15&   1.40&   6.69 &-    \\
mips009751 &-0.4550E-15 &-0.1140E-15 &-0.8730E-15&   1.40&   8.01 &-    \\
mips009834 &-0.3930E-14 &-0.1530E-14 &-0.6610E-14&   1.40&  10.78 &-    \\
mips010694 &-0.2060E-15 &-0.5030E-16 &-0.3930E-15&   1.40&   5.03 &-    \\
mips010818 &-0.4000E-15 &-0.1380E-15 &-0.6350E-15&   1.40&   7.75 &-    \\
mips013065 &-0.3700E-15 &-0.9570E-16 &-0.6880E-15&   1.40&   7.56 &-    \\
mips013589 &-0.4810E-15 &-0.1330E-15 &-0.8790E-15&   1.40&   8.96 &-    \\

\hline
\end{tabular}

Notes. --- Column~(1): MIPS $24\,\mu$m name. Column~(2): {\it Chandra}
full-band $0.5-8\,$keV flux or upper limit. Column~(3): {\it Chandra}
soft-band $0.5-2\,$keV flux or upper limit. Column~(4): {\it Chandra}
hard-band $2-8\,$keV flux or upper limit.  Negative values for the fluxes mean upper limits.
Column~(5): Photon index, for the sources not detected in all three {\it
  Chandra} bands, the photon  index is assumed to be $\Gamma=1.4$.
Column~(6): Off-axis angle in arcmin. Column~(7): If an ID is given,
cross-identification  with a previously published X-ray source. 
Source IDs starting with an 'A' are  from the Alexander et
al. (2003a) catalog, source ID G319 is from the 
Giacconi et al. (2002) catalog, and
source IDs starting with 'SUPP' are from the supplementary (unpublished data)
catalog (see text for details).
\end{table*}

\end{document}